\documentclass[pdflatex,sn-mathphys-num]{sn-jnl}% Math and Physical Sciences Numbered Reference Style
\usepackage{graphicx}%
\usepackage{multirow}%
\usepackage{amsmath,amssymb,amsfonts}%
\usepackage{amsthm}%
\usepackage{mathrsfs}%
\usepackage[title]{appendix}%
\usepackage{xcolor}%
\usepackage{textcomp}%
\usepackage{manyfoot}%
\usepackage{booktabs}%
\usepackage{algorithm}%
\usepackage{algorithmicx}%
\usepackage{algpseudocode}%
\usepackage{listings}%
\usepackage{mathtools}%

\usepackage{mymacros}

\theoremstyle{thmstyleone}%
\theoremstyle{thmstyletwo}%

\theoremstyle{thmstylethree}%

\begin{document}

\title[Active filament in shear flow]{Weakly nonlinear dynamics of a follower-force active filament in simple shear flow}

\author*{\begin{center}\fnm{Ory} \sur{Schnitzer} \\[1em]Department of Mathematics, Imperial College London, \\  London SW7 2AZ, United Kingdom\end{center}}

\abstract{We employ weakly nonlinear theory to investigate how an externally imposed simple shear flow affects the onset of spontaneous dynamics of an inertialess active filament deforming in Stokes flow. The filament, clamped to a wall at one end, is subjected to a compressive ``follower force'' applied tangentially at its free end. By extending my earlier weakly nonlinear analysis of the shear-free case (J.\ Fluid Mech., \textbf{1007} A65, 2025), we derive a generalized amplitude equation governing the near-onset dynamics under weak shear. Analysis of the amplitude equation shows that, besides inducing a steady deflection, the shear damps the filament’s intrinsic oscillations. This damping arises from a subtle nonlinear resonance between the shear and the intrinsic oscillations, scales quadratically with the shear rate, and is anisotropic---stronger in the flow direction than normal to it. Without shear, stable whirling states, where the filament tip traces a circular orbit in a plane parallel to the wall, and unstable planar-beating states are known to simultaneously emerge at a critical follower-force value, with circular whirling typically observed beyond this threshold. Shear breaks this degeneracy, driving a sequence of dynamical transitions: from circular to elliptical whirling, then to transverse beating (normal to the shear), and ultimately to steady deflection.}

%\keywords{Active filament, Stokes flow, weakly nonlinear analysis}



\maketitle

\let\thefootnote\relax\footnotetext{Email: o.schnitzer@imperial.ac.uk}

\section{Introduction}
\label{sec:intro}
Active filaments transform stored or ambient energy into deformation and motion. Of special interest is the possibility of such dynamics emerging \emph{spontaneously}, through instability and nonlinear symmetry breaking, rather than resulting from 
built-in asymmetry or detailed actuation. A fundamental model illustrating this paradigm consists of a slender filament clamped at one end to a wall and subjected to a compressive ``follower force'' applied tangentially at its free end. When the force exceeds a critical threshold, the filament buckles. Unlike classical Euler buckling, however, where the load remains vertical and leads to static deflection, a follower force gives rise to oscillatory motion. 

Early studies of the follower-force model---known in the structural mechanics literature as Beck's column---focused on inertial regimes \citep{Beck:52,Wood:69,Langthjem:00}. More recently, the model has been revisited in the context of biological microtubule or actin filaments, where the follower force mimics axial loading due to translocating molecular motors \citep{Bayly:16,De:17, Ling:18,Clarke:24}. In this context, inertia is typically neglected for both the filament and its surrounding viscous liquid. The deformation of the filament is thus governed by a balance between internal elastic stresses and external hydrodynamic loads, with the latter determined in the Stokes flow limit. While the follower-force model does not capture the full complexity of these biological systems, it nevertheless illustrates how molecular motors can give rise to rich nonlinear dynamics even without detailed biochemical regulation. 

The dynamics of a single follower-force active filament in the noninertial regime have been extensively studied. Initial-value simulations show that, in a force range just beyond the instability threshold, the filament typically settles into a three-dimensional ``whirling'' motion characterized by a locked shape steadily rotating about the vertical axis, with the tip tracing a circular orbit in a plane parallel to the wall \citep{Ling:18}. However, if the deformation is restricted to a plane (e.g., by the initial conditions), the filament instead exhibits ``beating'' oscillations \citep{De:17}. At higher force magnitudes, beating becomes stable while whirling loses stability; ultimately, the dynamics turn chaotic \citep{Ling:18}. 

Going beyond initial-value simulations, \citet{Clarke:24} employed computational dynamical-systems techniques to analyze the different dynamical states and their stability. They showed that beating and whirling emerge simultaneously via a double-Hopf bifurcation, with whirling near the onset of spontaneous motion consisting of specific phased superpositions of linear beating modes in nonparallel planes. Furthermore, they showed that near-onset beating is unstable under nonplanar perturbations, rationalizing the preference for whirling in that regime. They also elucidated the dynamical transitions observed at higher follower-force magnitudes. 

I have recently conducted a weakly nonlinear analysis of the follower-force model \citep{Schnitzer:25filament}. Using the method of multiple scales, I derived a nonlinear amplitude equation that captures the long-time dynamics of the filament near the instability threshold. Straightforward analysis of this equation furnished simple analytical expressions describing the beating and whirling states near the threshold, as well as their stability. 

In this paper, we extend that weakly nonlinear analysis to incorporate the effects of an externally imposed simple shear flow. We assume that the shear flow is weak enough that the induced steady deflection remains small relative to the filament length, yet strong enough to appreciably influence the near-onset dynamics. The present extension is motivated by the observation that active filaments often operate in fluid environments involving background flows. In particular, computational models of cytoplasmic streaming in cells, driven collectively by active-filament arrays interacting within a confined geometry, suggest that steady flows tend to stabilize spontaneous filament oscillations \citep{Stein:21,Dutta:24}. As we shall see, our minimal model of a single active filament in a prescribed background flow predicts a similar trend. 

The paper is structured as follows. In \S\ref{sec:formulation}, we formulate the problem. In \S\ref{sec:linear}, we conduct a linear small-deformation analysis, finding the deformation at the instability threshold as a superposition of steady deflection and periodic oscillations of the same  form as in the shear-free case \citep{Schnitzer:25filament}. In \S\ref{sec:WNT}, we analyze the long-time nonlinear dynamics near the instability threshold. This analysis employs a nonlinear amplitude equation that generalizes the one derived in \citep{Schnitzer:25filament} to include the effects of the externally imposed shear flow; the detailed derivation of the generalized amplitude equation via a multiple-scales weakly nonlinear expansion is provided in Appendix \ref{app:WNA}. In \S\ref{sec:conclusions}, we summarize our findings and discuss directions for future work. 

\section{Problem formulation}
\label{sec:formulation} 
We consider an inextensible elastic filament of length $L_*$ whose undeformed (reference) configuration is straight. The filament is clamped perpendicularly to a flat wall at one end and subjected to a tangential compressive force of magnitude $\mathcal{F}_*$ at the other. The filament is immersed in a fluid of viscosity $\eta_*$ and subject to a far-field simple shear flow parallel to the wall, of shear rate $\mathcal{G}_*$. The filament is modeled as a homogeneous Kirchhoff rod \citep{Landau:BookE} having a circular cross-section of radius $\kappa L_*$,  where $\kappa$ is a dimensionless slenderness parameter. The fluid is assumed to satisfy impermeability and no slip at the wall and on the filament surface. We neglect the inertia of both the filament and the fluid; this is easily justified assuming the small scales, low velocities and viscous fluids relevant to biological filaments. 

\begin{figure}
\begin{center}
\includegraphics[scale=0.45]{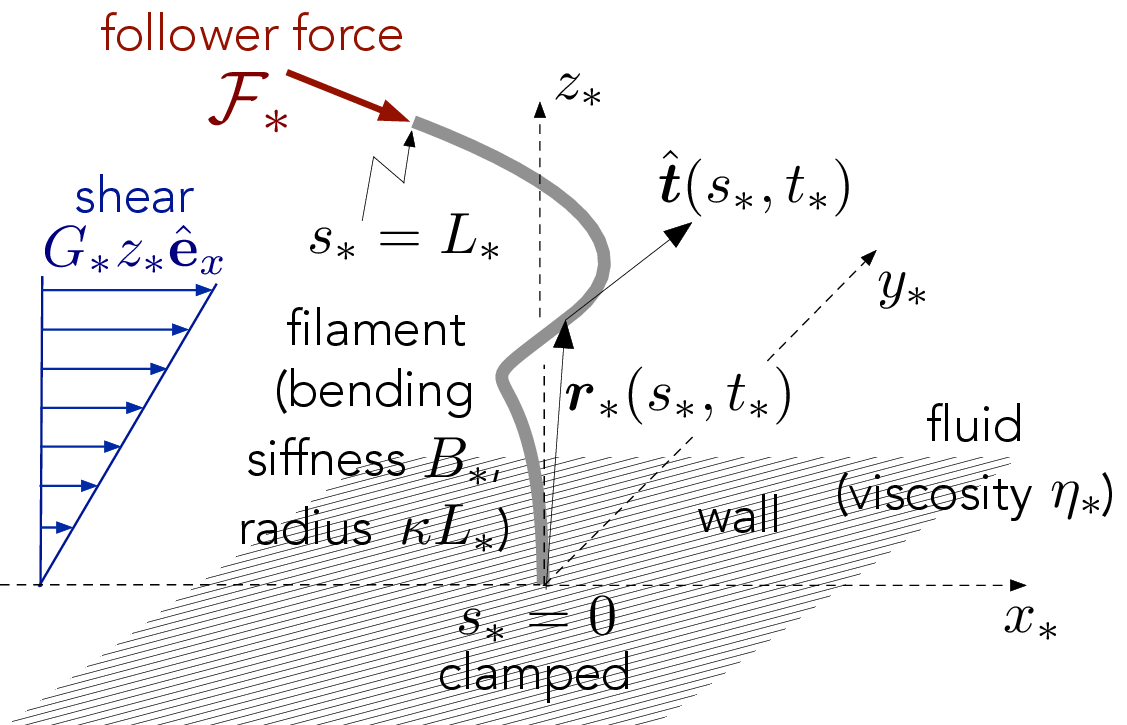}
\caption{Dimensional schematic of the problem.}
\label{fig:schematic}
\end{center}
\end{figure}

Let $\br_*(s_*,t_*)=x_*(s_*,t_*)\be_x+y_*(s_*,t_*)\be_y+z_*(s_*,t_*)\be_z$ be the filament centerline parameterized by the arc length $s_*$ measured from the clamping point and time $t_*$, with a tangent unit vector $\bt=\partial\br_*/\partial s_*$. Here $(x_*,y_*,z_*)$ denotes a Cartesian coordinate system whose origin is at the clamping point. The ``horizontal'' $x_*$ and $y_*$ axes run parallel to the wall, with the ``longitudinal'' direction $\be_x$ along the external shear flow. The ``vertical'' $z_*$ axis measures the distance from the wall, with $\be_z$ pointing into the fluid. The simple shear flow imposed in the far field is of the form $\mathcal{G}_*z_*\be_x$, which is consistent with the no-slip condition at the wall. A schematic of the problem is shown in Fig.~\ref{fig:schematic}. 

The deformation of the filament is governed by the force and moment balances
\refstepcounter{equation}
\label{force and moment balances dim}
$$
\pd{\bF_*}{s_*}+\bbf_*=\bzero, \qquad \pd{\bM_*}{s_*}+\bt\times \bF_*+\bm_*=\bzero,
\eqno{(\theequation \mathrm{a},\mathrm{b})}
$$ 
where $\bF_*(s_*,t_*)$ and $\bM_*(s_*,t_*)$ are the cross-sectionally averaged internal (elastic) force and moment, while $\bbf_*(s_*,t_*)$ and $\bm_*(s_*,t_*)$ are the  hydrodynamic force and moment acting on the filament per unit length due to Stokes flow at zero Reynolds number. 

Following \citet{De:17} and \citet{Ling:18}, we adopt resistive-force theory as a simplified description of the fluid-structure interaction that holds for Stokes flow, to leading order in the slender-filament limit $\kappa\ll1$ \citep{Graham:Book}. We accordingly approximate the hydrodynamic force density as
\begin{equation}\label{sbt dim}
\bbf_*=\frac{4\pi\eta_*}{\ln(1/\kappa)}\left(\tI-\frac{1}{2}\bt\bt\right)\bcdot \left(\mathcal{G}_*z_*\be_x-\pd{\br_*}{t_*}\right),
\end{equation}
where the rightmost parenthetical factor represents the relative fluid velocity incident on the filament. The local anisotropic drag law \eqref{sbt dim} is a crude approximation: the filament's interactions with itself and the wall are neglected despite the relative error being only logarithmically small in the slenderness, scaling as $1/\ln\kappa$. The hydrodynamic moment density $\bm_*$ is neglected as it is algebraically small in the slenderness. 
 
Since axial moments are absent and the filament is cross-sectionally isotropic (owing to its circular cross-sectional shape and homogeneous composition), the contribution to the internal moment due to twisting vanishes \citep{Landau:BookE}. The constitutive relation for the internal moment therefore reduces to 
\begin{equation}\label{constitutive dim}
\bM_*=B_*\bt\times\pd{\bt}{s_*},
\end{equation}
where the constant $B_*$ is the bending stiffness. 

Henceforth, we adopt a dimensionless convention where lengths are normalized by $L_*$, forces by $B_*/L_*^2$, moments by $B_*/L_*$ and time by $4\pi\eta_*(L_*^4/B_*)/\ln(1/\kappa)$. Dimensionless fields are denoted similarly to their dimensional counterparts, with the asterisk subscript omitted. 
The resulting dimensionless problem consists of the force and moment balances [cf.~\eqref{force and moment balances dim} and \eqref{sbt dim}]
\refstepcounter{equation}
$$
\label{force and moment balances}
\pd{\bF}{s}+\left(\tI-\frac{1}{2}\bt\bt\right)\bcdot \left(\mathcal{G} z\be_x-\pd{\br}{t}\right)=\bzero, \qquad \pd{\bM}{s}+\bt\times\bF=\bzero,
\eqno{(\theequation \mathrm{a},\mathrm{b})}
$$
where we have introduced the dimensionless shear rate 
\begin{equation}
\mathcal{G} = \frac{L_*^4\mathcal{G}_*}{B_*}\frac{4\pi\eta_*}{\ln(1/\kappa)};
\end{equation}
the geometric relations
\refstepcounter{equation}
$$
\label{geometric relations}
\pd{\br}{s}=\bt, \qquad \bt\bcdot\bt=1,
\eqno{(\theequation \mathrm{a},\mathrm{b})}
$$
which enforce inextensibility; 
the constitutive relation [cf.~\eqref{constitutive dim}]
\begin{equation}
\label{constitutive}
\bM=\bt\times\pd{\bt}{s};
\end{equation}
the clamping boundary conditions at the wall,
\refstepcounter{equation}
$$
\label{bcs wall}
\br=\bzero, \quad \bt=\be_z \quad \text{at} \quad s=0;
\eqno{(\theequation \mathrm{a},\mathrm{b})}
$$
and the tip conditions
\refstepcounter{equation}
$$
\label{bcs tip} 
\bF= -\mathcal{F}\bt, \quad \bM=\bzero \quad \text{at} \quad s=1,  \eqno{(\theequation \mathrm{a},\mathrm{b})}
$$
where we have introduced the dimensionless follower-force magnitude
\begin{equation}\label{parameter}
\mathcal{F}=\frac{\mathcal{F}_*L_*^2}{B_*}.
\end{equation}

\section{Linear theory}\label{sec:linear}
\subsection{Linearization under weak shear}
\label{ssec:linearproblem}
In the absence of the shear flow $(\mathcal{G}=0)$, the problem possesses the steady state
\refstepcounter{equation}
$$
\label{base state}
\br=s\be_z, \quad \bt=\be_z, \quad \bF=-\mathcal{F}\be_z, \quad \bM=\bzero,
\eqno{(\theequation \mathrm{a},\mathrm{b},\mathrm{c},\mathrm{d})}
$$
where the filament is vertical and under uniform compression by the follower force. In the weak-shear limit $\mathcal{G}\ll1$, we may consider small perturbations about  \eqref{base state},
\refstepcounter{equation}
$$
\br-s\be_z \approx \acute{\br}, \quad \bt-\be_z\approx \acute{\bbt}, \quad \bF+\mathcal{F}\be_z\approx \acute{\bF}, \quad \bM\approx \acute{\bM}.
\eqno{(\theequation \mathrm{a},\mathrm{b},\mathrm{c},\mathrm{d})}
$$
Upon substituting these perturbations into the governing equations \eqref{force and moment balances}--\eqref{bcs tip} and subsequently neglecting terms that are nonlinear in the perturbations, and those that are linear in the perturbations but also small in the shear, we find a linearized problem consisting of the differential equations
\begin{subequations}
\label{linearised eqs}
\begin{gather*}
\pd{\acute{\br}}{s}=\acute{\bbt}, \quad \pd{\acute{\bF}}{s}-\left(\tI-\frac{1}{2}\be_z\be_z\right)\bcdot\pd{\acute{\br}}{t}=-\mathcal{G}s\be_x, \tag{\theequation$\mathrm{a},\mathrm{b}$} \\ \pd{\acute{\bM}}{s}+\mathcal{F}\be_z\times\acute{\bbt}+\be_z\times\acute{\bF}=\bzero, \quad \acute{\bM}=\be_z\times\pd{\acute{\bbt}}{s}; 
\tag{\theequation$\mathrm{c},\mathrm{d}$}
\end{gather*}
\end{subequations}
the constraint
\begin{equation}
\label{linearised constraint}
\be_z\bcdot\acute{\bbt}=0;
\end{equation}
and the boundary conditions
\refstepcounter{equation}
$$
\label{linearised bcs}
\acute{\br}=\bzero, \quad \acute{\bbt}=\bzero \quad \text{at} \quad s=0; \qquad \acute{\bF}+\mathcal{F}\acute{\bbt}=\bzero, \quad \acute{\bM}=\bzero \quad \text{at} \quad s=1. 
\eqno{(\theequation \mathrm{a}\mathrm{-}\mathrm{d})}
$$

The constraint \eqref{linearised constraint} dictates that $\acute{\bbt}$ is horizontal. It then follows from the linearized geometric relation (\ref{linearised eqs}a) and the linearized clamping boundary condition (\ref{linearised bcs}a) that $\acute{\br}$ is also horizontal, and from the linearized constitutive relation (\ref{linearised eqs}d) that so is $\acute{\bM}$. Finally, the linearized force balance (\ref{linearised eqs}b) and the linearized tip condition (\ref{linearised bcs}c) show that $\acute{\bF}$ is horizontal as well. Hence, the linearized problem \eqref{linearised eqs}--\eqref{linearised bcs} governs horizontal vector fields. As a consequence, the term $(\tI-\be_z\be_z/2)\bcdot\partial\acute{\br}/\partial t$ appearing in (\ref{linearised eqs}b) reduces to $\partial\acute{\br}/\partial t$. 

\subsection{Particular solution: steady deflection}
\label{ssec:particular}
The linearized problem \eqref{linearised eqs}--\eqref{linearised bcs} is inhomogeneous owing to the shear-induced term on the right-hand side of the force balance (\ref{linearised eqs}b). We first construct a particular solution representing steady deflection of the filament, namely the leading-order positive-$\mathcal{G}$ continuation of the vertical steady state \eqref{base state} as $\mathcal{G}\to0$. Assuming a steady state and longitudinal deformation of the filament, it is straightforward to obtain this solution as
\begin{subequations}
\label{steady deflection}
\begin{gather*}
\acute{\br}=\mathcal{G} \ell(s)\be_x, \quad \acute{\bbt}=\mathcal{G}\ell'(s)\be_x, \tag{\theequation$\mathrm{a},\mathrm{b}$}\\ 
\acute{\bF}=\frac{1}{2}\mathcal{G}\left[1-s^2-2\mathcal{F}\ell'(1)\right]\be_x, \quad \acute{\bM}=\mathcal{G}\ell''(s)\be_y, \tag{\theequation$\mathrm{c},\mathrm{d}$}
\end{gather*}
\end{subequations}
where the deflection profile $\ell(s)$ satisfies the boundary-value problem 
\begin{equation}\label{steady bvp}
\ell''''+\mathcal{F}\ell''=s, \quad \ell(0)=\ell'(0)=\ell''(1)=\ell'''(1)=0.
\end{equation} 
(Henceforth, primes denote ordinary derivatives with respect to $s$.) Solving \eqref{steady bvp} gives
\begin{multline}\label{ell solution}
\ell(s)=\frac{1}{6\mathcal{F}}s^3-\frac{\cos \sqrt{\mathcal{F}}}{\mathcal{F}^2}(1+s)+\frac{1}{\mathcal{F}^2}\cos[\sqrt{\mathcal{F}}(1-s)] 
\\
+\frac{\sin\sqrt{\mathcal{F}}}{\mathcal{F}^{5/2}}(1-\mathcal{F}s)-\frac{1}{\mathcal{F}^{5/2}}\sin[\sqrt{\mathcal{F}}(1-s)].
\end{multline}

\begin{figure}[t]
\begin{center}
\includegraphics[scale=0.4,trim={2.5cm 0 0 0}]{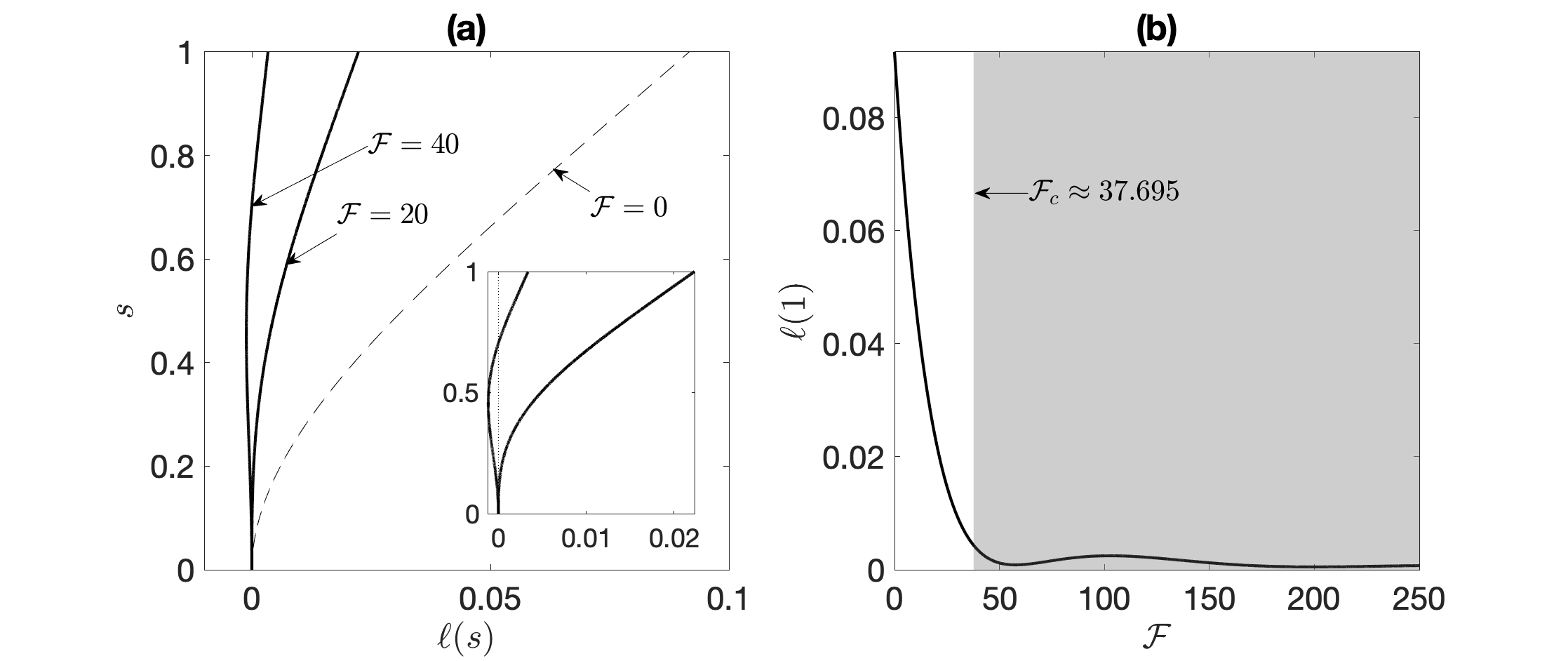}
\caption{Steady-state deflection under weak shear ($\mathcal{G}\ll1$): $\br\approx s\be_z + \mathcal{G}\ell(z)
\be_x$ (see \S\ref{ssec:particular}). \textbf{(a)} Deflection profile $\ell(s)$ for several values of $\mathcal{F}$. The inset shows the $\mathcal{F}\ne0$ profiles on a stretched scale. \textbf{(b)} Variation of the tip deflection $\ell(1)$ as a function of $\mathcal{F}$. The shaded region indicates the regime where the vertical steady state is unstable in the absence of shear; as explained in \S\ref{ssec:homogeneous}, this also approximates the regime where the steady-deflection state is unstable under weak shear.}
\label{fig:steadydeflection}
\end{center}
\end{figure}

This deflection profile depends on the follower force $\mathcal{F}$ as a parameter. For $\mathcal{F}\ll1$, \eqref{ell solution} gives $\ell(s)\sim s^2(s^3-10s+20)/120$, 
which agrees, to leading order in $\mathcal{G}$, with the profile derived by \citet{Kurzthaler:23} for a nonactive filament in simple shear flow. For $\mathcal{F}\gg1$, the deflection is small, $\ell(s)\sim s^3/(6\mathcal{F})$. Fig.~\ref{fig:steadydeflection} shows $\ell(s)$ for several values of $\mathcal{F}$, along with the corresponding tip deflection $\ell(1)$ as a function of $\mathcal{F}$. 
We see that steady deflection is diminished by the follower force. Intuitively, the shear acts to deflect the filament in the flow direction $\be_x$, resulting in the tangent at the tip satisfying $\bt\bcdot\be_x>0$. The follower force then has a negative component in that direction, opposing the shear. Since the shear-induced drag is distributed along the filament while the follower force is concentrated at the tip, the deflection profile becomes nonmonotone for sufficiently large $\mathcal{F}$, as depicted for $\mathcal{F}=40$. 

\subsection{Homogeneous problem: stability, beating and whirling} \label{ssec:homogeneous}
The general solution to the linearized weak-shear problem \eqref{linearised eqs}--\eqref{linearised bcs} is the sum of the steady deflection derived in \S\ref{ssec:particular} and a time-dependent solution to the homogeneous problem obtained by omitting the shear term in (\ref{linearised eqs}b). Consequently, to leading order in $\mathcal{G}$, the linear stability of the steady-deflection state is dictated by the same eigenvalue problem that governs the stability of the vertical steady state \eqref{base state} in the shear-free case. Below, we summarize key results from my earlier linear-stability analysis \citep{Schnitzer:25filament} of that homogeneous setup. As discussed therein, that analysis builds on those of \citet{De:17} and \citet{Ling:18}, and incorporates important insights from the computational study of \citet{Clarke:24}. 
 
The linear-stability analysis in \citep{Schnitzer:25filament} starts by assuming perturbations of the form 
\begin{equation}
\{\acute{\br},\acute{\bbt},\acute{\bF},\acute{\bM}\}=e^{\lambda t}\{\tilde{\br},\tilde{\bbt},\tilde{\bF},\tilde{\bM}\} + \text{c.c.},
\end{equation} 
where $\text{c.c.}$ stands for complex conjugate. This leads to an eigenvalue problem for the eigenfunctions $\{\tilde{\br},\tilde{\bbt},\tilde{\bF},\tilde{\bM}\}$ and growth-rate eigenvalues $\lambda$. Solving this problem reveals that the base state \eqref{base state} is unstable for $\mathcal{F}>\mathcal{F}_c\doteq37.695$, and that the onset of instability is oscillatory and degenerate: at the instability threshold $\mathcal{F}=\mathcal{F}_c$, two identical pairs of complex-conjugate eigenvalues cross the imaginary axis at $\lambda=\pm i\omega$, with $\omega\doteq 191.26$ being the angular frequency of neutrally stable oscillations (later referred to as the `instability frequency' or `natural frequency'). The most general neutral oscillation at the threshold is found as
\begin{equation}\label{linear homogeneous threshold}
\acute{\br}(s,t) = \bA e^{i\omega t} \varphi(s)+\text{c.c.},
\end{equation}
where $\bA$ is a horizontal complex-valued vector and $\varphi(s)$ is a complex-valued eigenfunction satisfying the boundary-value problem
\begin{equation}\label{phi def text}
\varphi''''+\mathcal{F}_c\varphi''+i\omega \varphi=0, \quad \varphi(0)=\varphi'(0)=\varphi''(1)=\varphi'''(1)=0,
\end{equation} 
along with a normalization condition chosen as $\varphi(1)=1$. The eigenfunction $\varphi(s)$ can be expressed in closed form up to the numerical constants $\mathcal{F}_c$ and $\omega$; for plots of the real and imaginary parts of $\varphi(s)$, see \citep{Schnitzer:25filament}. 
The corresponding solutions for the tangent, internal force and moment are $\acute{\bbt} = \bA e^{i\omega t}\varphi' + \text{c.c.}$, $\acute{\bF} = -\bA e^{i\omega t}(\varphi'''+\mathcal{F}_c\varphi') +\text{c.c.}$ and $\acute{\bM}= \be_z\times \bA e^{i\omega t}\varphi'' + \text{c.c.}$, respectively. 

To characterize the neutral oscillations \eqref{linear homogeneous threshold}, consider first the case where the real and imaginary parts of $\bA$ are parallel. In that case, we can write $\bA=B\hat{\bb}$, where $B$ is a complex constant and $\hat{\bb}$ a horizontal unit vector. We then see from \eqref{linear homogeneous threshold} that each point along the filament undergoes a time-harmonic oscillation along $\hat{\bb}$, centered about the vertical axis. Thus, the filament moves in the plane of $\hat{\bb}$ and $\be_z$. We shall refer to such planar motions as ``linear beating.'' Given the normalization $\varphi(1)=1$, the corresponding oscillation of the filament tip has a maximal displacement $2|B|$ and phase $\angle B$. 

\begin{figure}[t!]
\begin{center}
\includegraphics[scale=0.55,trim={1.6cm 0.5cm 0 0}]{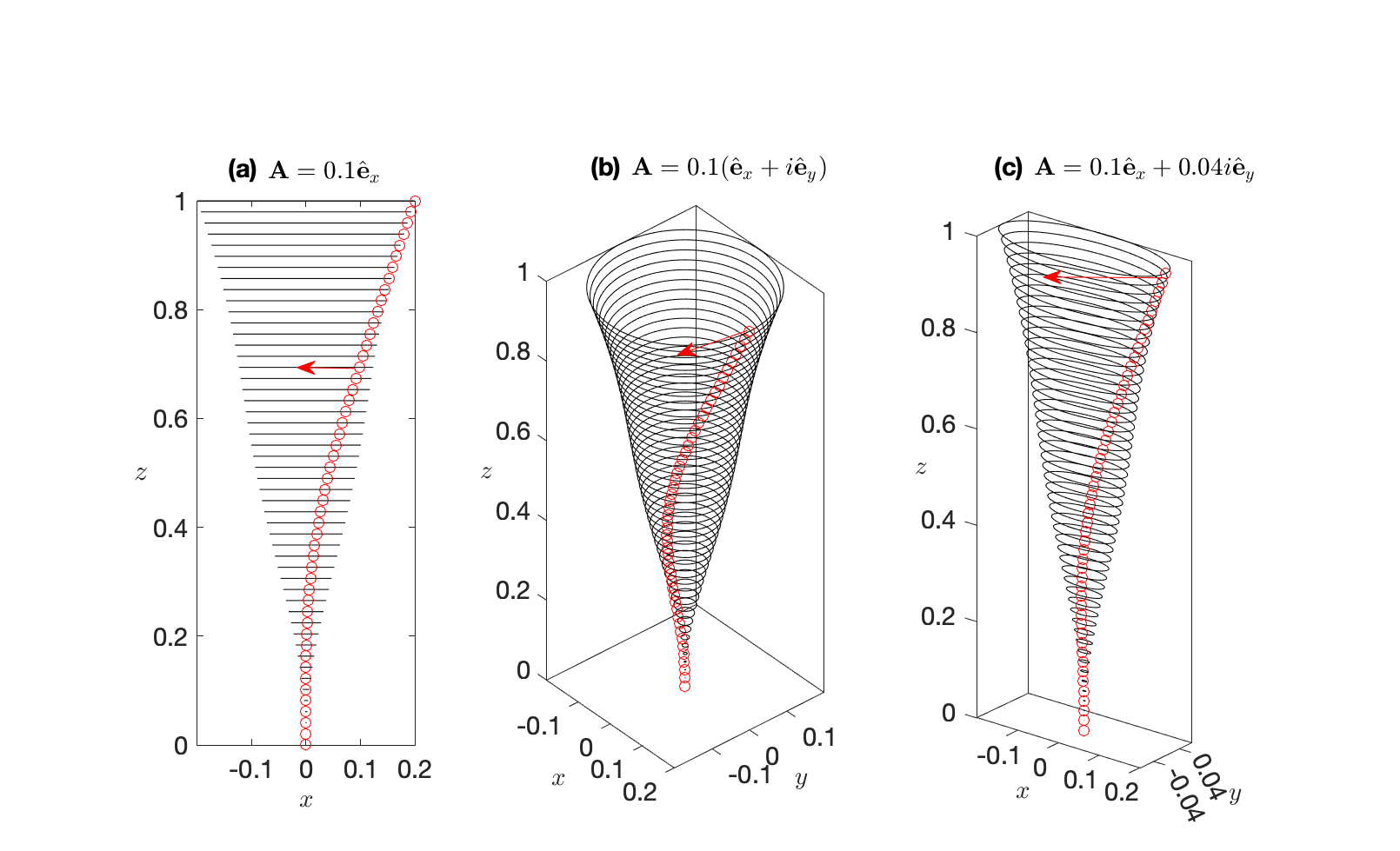}
\caption{The linear elliptical-whirling solutions \eqref{linear homogeneous threshold} depicted for three representative values of the complex-vector amplitude $\bA$. Each point along the centerline performs the same horizontal elliptical trajectory up to a phase difference (indicated by the circles) and a dilation. Case \textbf{(a)} is an example of linear planar beating, where the tip ellipse degenerates to a line. Case \textbf{(b)} is an example of linear circular whirling, where the tip ellipse is a circle. In case \textbf{(c)}, the tip trajectory is a noncircular ellipse.}
\label{fig:cones}
\end{center}
\end{figure}

Any amplitude $\bA$ can be expressed as $B_1\hat{\bb}_1+B_2\hat{\bb}_2$, where $B_1$ and $B_2$ are complex constants and $\hat{\bb}_1$ and $\hat{\bb}_2$ are nonparallel horizontal unit vectors. Thus, any motion of the form \eqref{linear homogeneous threshold} can be obtained via a phased superposition of, at most, two linear-beating motions in different planes. We shall refer to this general motion as ``linear elliptical whirling,'' since the tip $e^{i\omega t} \bA+\text{c.c.}$ traces a horizontal elliptical trajectory. In fact, all points along the centerline carry out the same elliptical trajectory up to a dilation $|\varphi(s)|$ and phase difference $\angle\varphi(s)$ relative to the tip trajectory. Fig.~\ref{fig:cones} depicts linear whirling oscillations for three representative values of $\bA$. 

Properties of the elliptical tip trajectory can be expressed in terms of $\bA$ (following \citet{Lindell:83}). To this end, we define the dot and cross products of two complex vectors $\ba$ and $\bb$ in terms of their counterparts for real vectors: $\ba\bcdot\bb=\ba_r\bcdot\bb_r-\ba_i\bcdot\bb_i+i(\ba_r\bcdot\bb_i+\ba_i\bcdot\bb_r)$,  $\ba\times\bb=\ba_r\times\bb_r-\ba_i\times\bb_i+i(\ba_r\times\bb_i+\ba_i\times\bb_r)$. We define the length of a complex vector based on the inner product $\langle \ba,\bb\rangle=\ba^*\bcdot\bb$, which differs from the dot product: $|\ba|=\sqrt{\ba^*\bcdot\ba}=\sqrt{|\ba_r|^2+|\ba_i|^2}$. Here, and throughout the paper, we use subscripts $r$ and $i$ to indicate real and imaginary parts. With these definitions, the lengths of the semi-major and semi-minor axes can be shown to be  $\sqrt{|\bA|^2+|\bA\times\bA^*|}\pm\sqrt{|\bA|^2-|\bA\times\bA^*|}$, respectively, where an asterisk superscript denotes the complex conjugate; the corresponding ellipse area is $2\pi|\bA\times\bA^*|$. The special case $\bA\times\bA^*=\bzero$ (i.e., $\bA_r$ and $\bA_i$ are parallel) corresponds to linear beating, where the ellipse degenerates to a line of length $4|\bA|$. The special case $\bA\bcdot\bA=0$ (i.e., $\bA_r$ and $\bA_i$ are orthogonal and of equal magnitude) corresponds to ``linear circular whirling,'' where the ellipse is a circle of radius $\sqrt{2}|\bA|$. For noncircular ellipses, the real and imaginary parts of $\bA/(\bA\bcdot\bA)^{1/2}$ point in the directions of the major and minor axes, respectively. Lastly, the direction of motion is indicated by $\mathrm{Im}(\bA\times\bA^*)$ according to the right-hand rule. 

\section{Weakly nonlinear theory}\label{sec:WNT}
\subsection{Separation of time scales near the instability threshold}\label{ssec:separation}
Let us summarize the linear weak-shear theory developed in the preceding section. (i) The instability threshold of the steady-deflection state is, to leading order as $\mathcal{G}\to0$, the same as for the vertical steady state in the shear-free setting: $\mathcal{F}=\mathcal{F}_c$. [We shall later show that the shear shifts this critical value by $O(\mathcal{G}^2)$.] (ii) For $\mathcal{F}<\mathcal{F}_c$, the solutions to the homogeneous linear problem decay exponentially, whereby the solution to the linear problem approaches the particular solution approximating the steady-deflection state. (iii) For $\mathcal{F}>\mathcal{F}_c$, there are exponentially growing solutions to the homogeneous linear problem; thus, the solution to the linear problem typically diverges away from steady deflection. (iv) At $\mathcal{F}=\mathcal{F}_c$, the solution approaches a superposition of steady deflection and linear elliptical whirling. 

Near the instability threshold, $\Delta\mathcal{F}=\mathcal{F}-\mathcal{F}_c\ll1$, the linear theory implies a similar superposition but with the whirling  slowly evolving over long times of order $1/\Delta\mathcal{F}$ [since the growth rates of the critical modes have an $O(\Delta\mathcal{F})$ real part]. Over long times, however, terms not included in the linearized weak-shear problem may become important. These include nonlinearities in the deformation, as well as terms that are both linear in the deformation and small in $\mathcal{G}$. We identify two long time scales associated with these neglected terms---in addition to the $1/\Delta\mathcal{F}$ time scale associated with the linearized dynamics. 

The first additional time scale follows from the analysis in \citep{Schnitzer:25filament} of the shear-free setting: $1/|\bA|^2$, with $|\bA|\ll1$ representing the small magnitude of the linear elliptical whirling. This time scale arises from a conventional weakly nonlinear mechanism where cubic nonlinear interactions of the linear whirling with itself result in $O(|\bA|^3)$ forcing terms that include ``first harmonics,'' i.e., time-harmonic terms of angular frequency $\omega$. These forcing terms, in turn, resonantly excite $O(|\bA|^3t)$ perturbations, which become comparable to the linear oscillation over $O(1/|\bA|^2)$ times. 

The second additional time scale, $\mathcal{G}^{-2}$, originates from a subtle nonlinear resonance between the weak shear flow and the filament's intrinsic oscillations. (A \emph{steady} shear flow cannot possibly resonate linearly with these time-harmonic oscillations.) Naively, a resonance may be thought to arise from forcing terms formed of quadratic products of the linear elliptical whirling  and the shear flow, giving $O(\mathcal{G}|\bA|)$ first harmonics; but owing to the form of the governing equations \eqref{force and moment balances} and \eqref{constitutive}, such products are oriented vertically and therefore fail to excite the horizontal whirling modes. The next-strongest products that involve the shear and generate first harmonics are cubic: linear in the oscillation and quadratic in the shear flow. Unlike the stronger quadratic products, such cubic products are horizontal, and may therefore resonate with the intrinsic oscillations. This would result in $O(\mathcal{G}^2|\bA|t)$ perturbations, becoming comparable to the linear oscillation over $O(\mathcal{G}^{-2})$ times.

\subsection{Amplitude-equation model}
\label{ssec:amplitude}
The above considerations motivate a multiple-scales weakly nonlinear analysis in the near-threshold and weak-shear distinguished limit where $\mathcal{G}=O(\Delta\mathcal{F}^{1/2})$ as $\Delta\mathcal{F}\to0$. Assuming that weak nonlinearity saturates the linearly unstable growth above the threshold, our scaling arguments imply, in that distinguished limit, oscillation amplitudes $|\bA|=O(\Delta\mathcal{F}^{1/2})$---comparable to the $O(\mathcal{G})$ steady deflection induced by the shear flow. Having identified these scalings, such an analysis becomes a direct, though  technically demanding, extension of my analysis of the shear-free problem \citep{Schnitzer:25filament}. The details of this derivation are given in the Appendix.

The weakly nonlinear analysis confirms that the deformation can be approximated as a superposition of steady deflection and linear elliptical whirling,
\begin{equation}\label{near onset with shear}
\br(s,t)- s\be_z =  \mathcal{G}\ell_c(s)\be_x + \{e^{i\omega t} \varphi(s)\bA(t)+\text{c.c.}\},
\end{equation}
with $\ell_c(s)$ denoting the steady deflection profile $\ell(s)$ evaluated at $\mathcal{F}=\mathcal{F}_c$ [cf.~\eqref{ell solution}], and the complex-vector amplitude $\bA(t)$ evolving according to the nonlinear amplitude equation
\begin{equation}\label{amplitude shear}
\frac{d\bA}{dt}=\alpha\bA^*\bA\bcdot\bA+\beta\bA\bA^*\bcdot\bA+(\gamma\Delta\mathcal{F}
-\xi^{\parallel}\mathcal{G}^2)\be_x\be_x\bcdot\bA
+(\gamma\Delta\mathcal{F}-\xi^{\perp}\mathcal{G}^2)\be_y\be_y\bcdot\bA, 
\end{equation}
in which 
\begin{subequations}
\label{coefficients}
\begin{gather*}
\alpha \doteq -451.038 -i327.599, \quad 
 \beta\doteq -517.974+i353.952,  \tag{\theequation$\mathrm{a},\mathrm{b}$} \\ 
  \gamma\doteq 7.34480+i5.34285, \tag{\theequation$\mathrm{c}$} \\
\xi^\parallel\doteq 0.0493226+i0.0172294, \quad \xi^{\perp}\doteq0.0181937-i0.0040939. \tag{\theequation$\mathrm{d},\mathrm{e}$}
\end{gather*}
\end{subequations}
The complex coefficients \eqref{coefficients} are defined in the Appendix as quadratures involving the eigenfunction $\varphi(s)$ [cf.~\eqref{phi def text}]; an adjoint eigenfunction; and, for $\xi^{\parallel}$ and $\xi^{\perp}$, also the steady-deflection profile $\ell_c(s)$. A Mathematica notebook performing these quadratures is provided as Supplementary Information, Part I. Below, we make several remarks regarding the above amplitude-equation model.  

\medskip
\noindent\textbf{Time scales.} Balancing the time derivative in \eqref{amplitude shear} against different terms on the right-hand side of that equation recovers the three long time scales (relative to the instability period $2\pi/\omega$) discussed in \S\ref{ssec:separation}: $1/\Delta\mathcal{F}$, representing slow linear growth or decay near the threshold; $1/|\bA|^2$, representing weak nonlinear effects under small deformations; and $1/\mathcal{G}^2$, representing the nonlinear coupling with the shear flow. The derivation of the amplitude-equation model assumes that these three long time scales are comparable.

\medskip
\noindent\textbf{Factorizations.} For $\mathcal{G}\ne0$, $\mathcal{G}$ can be factored out by scaling $\bA$ by $\mathcal{G}$, $\Delta\mathcal{F}$ by $\mathcal{G}^2$, and $t$ by $1/\mathcal{G}^2$. Alternatively, for $\Delta\mathcal{F}\ne0$, $\Delta\mathcal{F}$ can be factored out by scaling $\bA$ and $\mathcal{G}$ by $|\Delta\mathcal{F}|^{1/2}$, and $t$ by $|\Delta\mathcal{F}|^{-1}$. It follows that, up to rescalings of the amplitude and time, the dynamics are governed by a single control parameter:
\begin{equation}
\frac{\Delta\mathcal{F}}{\mathcal{G}^2}.
\end{equation}
%\Delta\mathcal{F}/\mathcal{G}^2$.

\medskip
\noindent\textbf{Effective anisotropic damping.} The imposed shear flow manifests in the amplitude equation \eqref{amplitude shear} as homogeneous terms---linear in the amplitude and quadratic in the shear rate---rather than inhomogeneous forcing terms as one might expect for an external forcing. This aligns with our earlier intuition that the weak shear influences the dynamics via a nonlinear resonance involving the natural modes linearly and the shear flow quadratically. We interpret the shear-induced homogeneous terms as a form of linear damping, with resistance coefficients that scale quadratically with the shear rate---the resistance in the longitudinal direction $\be_x$ being nearly triple that in the transverse direction $\be_y$ [cf.~\eqref{coefficients}]. (Since $\xi^{\parallel}$ and $\xi^{\perp}$ are complex-valued, this damping is oscillatory in nature.) 

\medskip
\noindent\textbf{Shear-free case.} 
Setting $\mathcal{G}=0$ recovers the amplitude equation derived in \citep{Schnitzer:25filament}. (Note a minor difference in notation: the complex-vector amplitude is denoted there by $\tilde{\bA}$, reserving $\bA$ for a rescaled amplitude.) We will review the dynamics predicted in this shear-free limit in \S\ref{ssec:summary}. 

\medskip
\noindent\textbf{Alternative formulations.} 
The amplitude equation \eqref{amplitude shear} represents a four-dimensional homogeneous and autonomous dynamical system. It can be written as a pair of real-vector equations by decomposing the complex-vector amplitude into its real and imaginary parts: $\bA(t)=\bA_r(t)+i\bA_i(t)$. Alternatively, it can be written as a pair of coupled complex-scalar amplitude equations by decomposing the complex-vector amplitude as $\bA(t)=A_x(t)\be_x+A_y(t)\be_y$, where $A_x(t)$ and $A_y(t)$ are complex Cartesian components. It can also be written as a real dynamical system by further decomposing these Cartesian components into their real and imaginary parts, e.g., $A_x(t)=A_{x,r}(t)+iA_{x,i}(t)$. We shall employ analogous decompositions for other complex vectors, denoted similarly. 

\medskip
\noindent\textbf{Symmetries.} 
As a consequence of the symmetry of the system about the $xz$ plane, the amplitude equation \eqref{amplitude shear} is invariant under the transformation $A_y(t)\mapsto -A_y(t)$; hence, if $\bA(t)={A}_x(t)\be_x+{A}_y(t)\be_y$ is a solution, then so is ${\bA}(t)={A}_x(t)\be_x-{A}_y(t)\be_y$. This corresponds to a reflection about the $xz$ symmetry plane. Furthermore, the amplitude equation is invariant under $\bA\mapsto e^{i\phi}\bA$, where $\phi$ is any constant real phase---a transformation representing a shift of the fast time. In particular, if $\bA(t)$ is a solution, then so is $-\bA(t)$. By composing the latter transformation with $A_y(t)\mapsto -A_y(t)$, it follows that the amplitude equation is also invariant under $A_x(t)\mapsto -A_x(t)$, though this does not represent an independent geometric symmetry. 

\medskip
\noindent\textbf{Invariance.} 
If $A_y(0)=0$, then $A_y(t)\equiv 0$ for $t>0$; and if $A_x(0)=0$, then $A_x(t)\equiv 0$ for $t>0$. Thus, purely longitudinal and purely transverse oscillations---superimposed on the steady longitudinal deflection---constitute invariant subspaces of the dynamics. In the shear-free setting, planar oscillations along any horizontal axis are invariant. 

\medskip
\noindent\textbf{Trivial steady state: steady deflection.} 
Since the amplitude equation \eqref{amplitude shear} is homogeneous, the trivial solution $\bA(t)\equiv\bzero$ always exists. From \eqref{near onset with shear}, this steady solution corresponds to the steady-deflection state. In fact, it follows from the phase symmetry described above that there are no other steady states (except, potentially, at isolated values of the control parameter $\Delta\mathcal{F}/\mathcal{G}^2$). This is because setting $d\bA/dt=\bzero$ in \eqref{amplitude shear} yields a system of 4 real algebraic equations, whereas the phase symmetry allows us to choose one of the four components of $\bA$ at will. 

\medskip
\noindent\textbf{Quasi-steady solutions: periodic states.} 
Periodic states are represented by quasi-steady solutions of the amplitude equation, 
\begin{equation}\label{quasi}
\bA=e^{i\nu t}\bar{\bA},
\end{equation}
where $\bar\bA$ is a time-independent horizontal complex vector and the time-harmonic prefactor represents a small perturbation of the angular frequency from $\omega$ to $\omega+\nu$. (The frequency shift $\nu$ is necessarily small given the slow timescales implied by the amplitude equation.) Substitution of the ansatz \eqref{quasi} into the amplitude equation \eqref{amplitude shear} yields a time-independent equation governing $\bar{\bA}$ and $\nu$,
\begin{equation}\label{amplitude steady}
\alpha\bar{\bA}^*\bar{\bA}\bcdot\bar{\bA}+\beta\bar{\bA}\bar{\bA}^*\bcdot\bar{\bA}+(\gamma\Delta\mathcal{F}-\xi^{\perp}\mathcal{G}^2-i\nu)\bar{\bA}- (\xi^{\parallel}-\xi^{\perp})\mathcal{G}^2\be_x\be_x\bcdot \bar{\bA}=\bzero.
\end{equation}
As a consequence of the continuous phase symmetry discussed above, quasi-steady solutions can only exist as a continuous family parameterized by an arbitrary real phase. 

\medskip
\noindent\textbf{Stability.} 
The stability of a quasi-steady solution \eqref{quasi} (generally representing a periodic state) can be analyzed by assuming infinitesimal perturbations of the form 
\begin{equation}\label{perturbation from beating}
\bA(t)- e^{i\nu t}\bar\bA \approx e^{i\nu t}\ba(t),
\end{equation} 
where factoring out $e^{i\nu t}$ from the perturbation leads to a linearized amplitude equation with time-independent coefficients,
\begin{multline}\label{linearised about quasisteady}
\frac{d\ba}{dt}=2\alpha\bar{\bA}^*\bar{\bA}\bcdot\ba+\alpha\ba^*\bar{\bA}\bcdot\bar{\bA}+\beta\bar{\bA}\bar{\bA}^*\bcdot\ba+\beta\bar{\bA}\ba^*\bcdot\bar{\bA}+(\beta|\bar{\bA}|^2-i\nu)\ba\\
+(\gamma\Delta\mathcal{F}-\xi^{\parallel}\mathcal{G}^2)\be_x\be_x\bcdot\ba+(\gamma\Delta\mathcal{F}-\xi^{\perp}\mathcal{G}^2)\be_y\be_y\bcdot\ba.
\end{multline}
The stability of a steady solution $\bA(t)=\bar{\bA}$ (corresponding to a true steady state) can be analyzed by setting $\nu=0$ in \eqref{linearised about quasisteady}. 

Written in terms of the real and imaginary parts of the Cartesian components of $\ba(t)$, \eqref{linearised about quasisteady} yields a four-dimensional linear, homogeneous and autonomous real dynamical system. Thus, the stability of both steady and quasi-steady solutions can be assessed by looking for exponential solutions of the form 
\begin{equation}
\label{exp ansatz}
\begin{pmatrix} a_{x,r} \\a_{x,i} \\ a_{y,r} \\ a_{y,i}\end{pmatrix}  = e^{i\sigma t}\begin{pmatrix} \tilde{a}_{x,r} \\\tilde{a}_{x,i} \\ \tilde{a}_{y,r} \\ \tilde{a}_{y,i}\end{pmatrix} +\text{c.c}, 
\end{equation}
leading to a matrix eigenvalue problem for the growth rates $\sigma$ and eigenvectors $\{\tilde{a}_{x,r}, \tilde{a}_{x,i}, \tilde{a}_{y,r}, \tilde{a}_{y,i}\}^T$, both of which are in general complex valued. Thus, within the framework of the weakly nonlinear theory, determining the linear stability of a periodic state reduces to a conventional modal analysis, obviating the need for a Floquet analysis. 

A simplification arises from the fact that the stability of a quasi-steady solution is independent of its phase. Indeed, the linearized amplitude equation \eqref{linearised about quasisteady} is invariant under the simultaneous transformations $\bar{\bA}\mapsto e^{i\vartheta}\bar{\bA}$ and $\ba\mapsto e^{i\vartheta}\ba $, where $\vartheta$ is an arbitrary real constant. Consequently, we may, without loss of generality, evaluate the stability of a family of quasi-steady states by choosing the phase at will. Furthermore, it is readily verified using the steady amplitude equation \eqref{amplitude steady}   that the perturbation $\ba=i\bar{\bA}$---corresponding to an infinitesimal phase shift---is always a solution to the linearized amplitude equation \eqref{linearised about quasisteady}, with $d\ba/dt=0$. For any nontrivial quasi-steady solution ($\bar{\bA}\ne\bzero$), this so-called ``Goldstone'' \citep{Chaikin_Lubensky_1995} mode is neutrally stable, corresponding to a zero growth rate $\sigma=0$. 

\subsection{Shear-free dynamics (overview)}\label{ssec:summary}
Before addressing the effect of the shear flow, it is instructive to first summarize the predictions of the amplitude equation \eqref{amplitude shear} in the shear-free case $\mathcal{G}=0$ previously analyzed in \citep{Schnitzer:25filament}. 

In that case, the trivial solution $\bA=\bzero$ corresponds to the vertical steady state \eqref{base state}. For this solution, the linearized amplitude equation \eqref{linearised about quasisteady} reduces to 
\begin{equation}
\frac{d\ba}{dt}= \gamma \Delta\mathcal{F}\ba,
\end{equation}
with the general solution 
\begin{equation}
\ba(t)=\ba(0)e^{\gamma\Delta\mathcal{F}t}.
\end{equation}
The corresponding filament deformation follows from \eqref{near onset with shear}:
\begin{equation}
\br(s,t)-s\be_z \approx \ba(0)e^{\gamma_r\Delta\mathcal{F}t}e^{i(\omega+\gamma_i\Delta\mathcal{F})t}\varphi(s)+\text{c.c.}
\end{equation}
Hence, infinitesimal deformations near the threshold consist of linear elliptical whirling with a time-dependent dilation $\exp(\gamma_r\Delta\mathcal{F}t)$ and an angular-frequency shift from $\omega$ to $\omega+\gamma_i\Delta\mathcal{F}$. Since $\gamma_r>0$, we conclude that the vertical steady state is oscillatory stable for $\Delta\mathcal{F}<0$ and oscillatory unstable for $\Delta\mathcal{F}>0$. Since $\ba(0)$ is an arbitrary complex vector in the horizontal plane, it represents the superposition of oscillations in two independent planes normal to the wall. Hence, the oscillatory onset of the instability at the threshold $\Delta\mathcal{F}=0$ occurs via a double-Hopf bifurcation, in accordance with the linear-stability analysis of the full shear-free problem. (A nonlinear analysis of the shear-free amplitude equation at the threshold shows that perturbations also decay in that case, albeit algebraically with time rather than exponentially.) 

Periodic states, represented by quasi-steady amplitudes of the form \eqref{quasi}, are found by solving the time-independent amplitude equation \eqref{amplitude steady}. Two distinct families of periodic states are found to concurrently emerge at the threshold and exist for $\Delta\mathcal{F}>0$. The first family consists of beating states, 
\begin{subequations}
\label{beating free}
\begin{gather}
\bar{\bA}=\hat{\bb}e^{i\vartheta}\sqrt{\frac{\gamma_r\Delta\mathcal{F}}{-(\alpha_r+\beta_r)}}, \quad \nu=\Delta\mathcal{F}\left(\gamma_i-\gamma_r\frac{\alpha_i+\beta_i}{\alpha_r+\beta_r}\right),
\tag{\theequation a,b}
\end{gather}
\end{subequations}
parameterized by an arbitrary real phase $\vartheta$ and the horizontal unit vector $\hat{\bb}$ defining the plane of beating. The second family consists of circular-whirling states, 
\begin{subequations}
\label{whirling free}
\begin{gather}
\bar\bA = e^{i\vartheta}(\be_x\pm i\be_y)\sqrt{\frac{\gamma_r\Delta\mathcal{F}}{-2\beta_r}}, \quad
\nu=\Delta\mathcal{F}\left(\gamma_i-\gamma_r\frac{\beta_i}{\beta_r}\right), \tag{\theequation a,b}
\end{gather}
\end{subequations}
parameterized by an arbitrary real phase $\vartheta$ and a sign choice that determines the direction of travel: a plus (minus) sign indicates clockwise (counter-clockwise) motion when viewed from above the wall. [Note that the vector factor $e^{i\vartheta}(\be_x\pm i\be_y)$ used here is algebraically equivalent, up to a redefinition of the arbitrary phase, to the geometric form $\hat{\boldsymbol{\vartheta}}\pm i \be_z\times \hat{\boldsymbol{\vartheta}}$ used in \citep{Schnitzer:25filament}, in which $\hat{\boldsymbol{\vartheta}}$ is an arbitrary horizontal unit vector.] In accordance with the scalings stated in the previous subsection, the oscillation amplitudes $|\bar{\bA}|$ scale with $|\Delta\mathcal{F}|^{1/2}$ while the angular-frequency corrections $\nu$ scale with $|\Delta\mathcal{F}|$, the latter corresponding to a slow modulation of the neutral linear elliptical whirling on the long time scale $1/|\Delta\mathcal{F}|$. 

Analysis of the linearized amplitude equation \eqref{linearised about quasisteady}, with $\bar{\bA}$ given by the above quasi-steady solutions, shows that the circular-whirling states are stable (up to the neutral mode associated with the arbitrary phase). In contrast, the beating states are stable under perturbations within  the plane of beating (up to a neutral mode associated with the arbitrary phase), but unstable under nonplanar perturbations; accordingly, the beating states are unstable. 
\begin{figure}
\begin{center}
\includegraphics[scale=0.5]{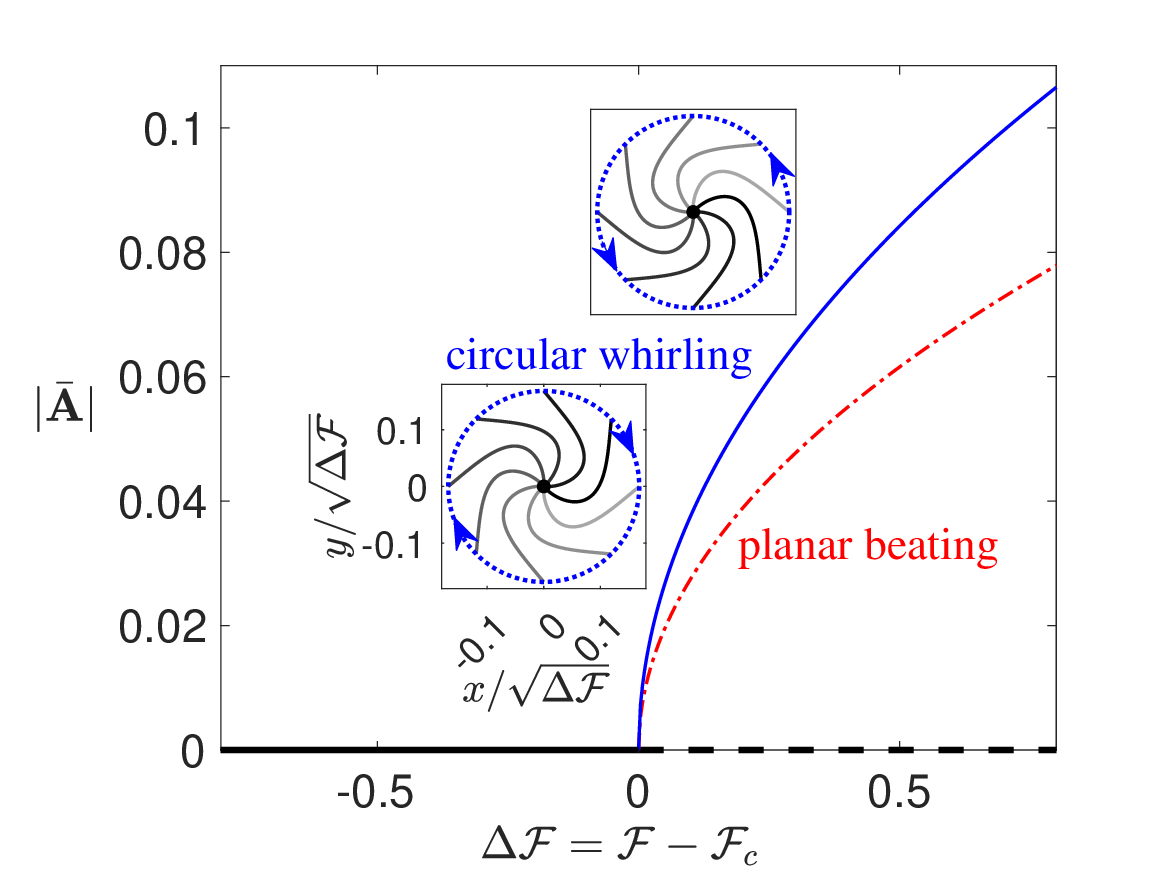}
\caption{Bifurcation diagram predicted by the weakly nonlinear theory in the shear-free case. Solid: stable or neutrally stable states. Dash-dotted: unstable, but neutrally stable under parallel perturbations. Dashed: unstable. The insets depict the clockwise and counter-clockwise circular-whirling states, with the greyscale solid curves showing horizontal projections of the filament centerline at equally spaced times over the period and the blue dotted circles tracing the tip orbits.}
\label{fig:noshear}
\end{center}
\end{figure}

Fig.~\ref{fig:noshear} shows the bifurcation diagram predicted by the weakly nonlinear theory in the shear-free scenario, with insets depicting the circular-whirling states. For $\Delta\mathcal{F}<0$, oscillations are damped. For $\Delta\mathcal{F}>0$, the filament typically settles into circular whirling, with the phase and direction of motion determined by the initial conditions; if, however, the initial conditions are precisely planar, i.e., $\bA_r(0)\parallel\bA_i(0)$, or $\bA(0)\times\bA^*(0)=\bzero$, the filament instead settles into beating. In \citep{Schnitzer:25filament}, I demonstrated these predictions via initial-value simulations of the shear-free amplitude equation and validated them against numerical simulations of the full problem (contributed by Eric Keaveny).

\section{Dynamics under weak shear}\label{sec:sheardynamics}
We now turn to the analysis of the amplitude equation \eqref{amplitude shear} including the anisotropic-damping terms arising from the imposed shear flow. 

\subsection{Stability of steady deflection: pair of Hopf bifurcations}
\label{ssec:stabilitydeflected}
For $\mathcal{G}\ne0$, the trivial solution $\bA(t)\equiv\bzero$ represents a steady-deflection state [cf.~\eqref{near onset with shear}]. Its stability is governed by the linearized amplitude equation \eqref{linearised about quasisteady} with $\bar{\bA}=\bzero$ and $\nu=0$, 
\begin{equation}\label{linear about steady deflection}
\frac{d\ba}{dt}=(\gamma\Delta\mathcal{F}-\xi^{\parallel}\mathcal{G}^2)\be_x\be_x\bcdot\ba+(\gamma\Delta\mathcal{F}-\xi^{\perp}\mathcal{G}^2)\be_y\be_y\bcdot\ba.
\end{equation}
To analyze this equation, we employ the Cartesian decomposition $\ba(t)=a_x(t)\be_x+a_y(t)\be_y$, with the $a_x(t)$ and $a_y(t)$ components representing longitudinal and transverse linear beating, respectively. In terms of these components, \eqref{linear about steady deflection} reduces to a pair of \emph{decoupled} differential equations,
\refstepcounter{equation}
$$
\frac{da_x}{dt}=(\gamma\Delta\mathcal{F}-\xi^{\parallel}\mathcal{G}^2)a_x, \qquad \frac{da_y}{dt}=(\gamma\Delta\mathcal{F}-\xi^{\perp}\mathcal{G}^2)a_y.
\eqno{(\theequation \mathrm{a},\mathrm{b})}
$$
Their general solutions are 
\refstepcounter{equation}
$$
a_x(t)=a_x(0)e^{(\gamma\Delta\mathcal{F}-\xi^{\parallel}\mathcal{G}^2)t}, \qquad 
a_y(t)=a_y(0)e^{(\gamma\Delta\mathcal{F}-\xi^{\perp}\mathcal{G}^2)t}.
\eqno{(\theequation \mathrm{a},\mathrm{b})}
$$ 
The corresponding linear approximation for the deformation follows from \eqref{near onset with shear}, 
\begin{multline}
\br(s,t)-s\be_z\approx  \be_xa_x(0)e^{(\gamma_r\Delta\mathcal{F}-\xi_r^{\parallel}\mathcal{G}^2)t}e^{i(\omega+\gamma_i\Delta\mathcal{F}-\xi_i^{\parallel}\mathcal{G}^2)t}\varphi(s) \\
+\be_ya_y(0)e^{(\gamma_r\Delta\mathcal{F}-\xi_r^{\perp}\mathcal{G}^2)t}e^{i(\omega+\gamma_i\Delta\mathcal{F}-\xi_i^{\perp}\mathcal{G}^2)t}\varphi(s)+\text{c.c.}
\end{multline}
It shows that the double-Hopf bifurcation found in the shear-free case at $\Delta\mathcal{F}=0$ splits into a pair of Hopf bifurcations at $\Delta\mathcal{F}$ values given by
\refstepcounter{equation}
$$
\label{thresholds}
\Delta\mathcal{F}^{\parallel} = \frac{\xi_r^{\parallel}}{\gamma_r}\mathcal{G}^2, \qquad \Delta\mathcal{F}^{\perp} = \frac{\xi_r^\perp}{\gamma_r}\mathcal{G}^2,
\eqno{(\theequation \mathrm{a},\mathrm{b})}
$$
respectively corresponding to the thresholds for longitudinal and transverse linear beating. 
Since $0<\xi_r^{\perp}<\xi_r^{\parallel}$, we have that $0<\Delta\mathcal{F}^{\perp}<\Delta\mathcal{F}^{\parallel}$. Hence, the instability threshold of the steady state is postponed to higher follower force relative to the shear-free case, from $\Delta\mathcal{F}=0$ to $\Delta\mathcal{F}=\Delta\mathcal{F}^{\perp}$. In the interval $\Delta\mathcal{F}^{\perp}<\Delta\mathcal{F}<\Delta\mathcal{F}^{\parallel}$, the steady state remains stable under longitudinal perturbations, thus stable in the special case where the dynamics are restricted to the $xz$ plane.

\subsection{Longitudinal and transverse beating states}\label{ssec:beatingstates}
The Hopf-bifurcation splitting described in the preceding subsection suggests that, with shear flow, whirling-like states---formed by superposing longitudinal and transverse linear beating---can no longer emerge \emph{directly} from the steady state. This motivates looking for planar beating states bifurcating from the thresholds \eqref{thresholds}.
 
Planar beating states correspond to quasi-steady solutions of the form \eqref{quasi}, with 
\begin{equation}\label{beating form}
\bar\bA=Be^{i\vartheta}\hat{\bb},
\end{equation} 
$\hat{\bb}$ being a horizontal unit vector, $\vartheta$ a real-valued phase and $B=|\bar{\bA}|>0$. With this ansatz, the steady amplitude equation \eqref{amplitude steady} becomes 
\begin{equation}\label{beating equation}
[(\alpha+\beta)B^2+\gamma\Delta\mathcal{F}-\xi^{\perp}\mathcal{G}^2-i\nu]\hat\bb
-(\xi^{\parallel}-\xi^{\perp})\mathcal{G}^2\be_x\be_x\bcdot\hat\bb=\bzero.
\end{equation}
Crossing both sides of this equation by $\hat\bb$ gives $(\be_x\times\hat\bb)(\be_x\bcdot\hat\bb)=\bzero$, so that either $\hat\bb=\be_x$, corresponding to longitudinal beating, or $\hat{\bb}=\be_y$, corresponding to transverse beating. 

In the longitudinal case, \eqref{beating equation} reduces to 
\begin{equation}\label{beating equation long}
(\alpha+\beta)B^2+\gamma\Delta\mathcal{F}
-\xi^{\parallel}\mathcal{G}^2-i\nu=0.
\end{equation}
We separate \eqref{beating equation long} into real and imaginary parts and solve the resulting pair of real equations for $B=B^{\parallel}$ and $\nu=\nu^{\parallel}$. Using (\ref{thresholds}a), we find solutions for $\Delta\mathcal{F}>\Delta\mathcal{F}^{\parallel}$, given by  
\begin{subequations}
\label{mag nu long}
\begin{align}
B^{\parallel}&=\sqrt{\frac{\gamma_r(\Delta\mathcal{F}-\Delta\mathcal{F}^\parallel)}{-(\alpha_r+\beta_r)}}, \\ 
\nu^{\parallel}&=
\left(\gamma_i-\frac{\alpha_i+\beta_i}{\alpha_r+\beta_r}\gamma_r\right)\Delta\mathcal{F}+\frac{\alpha_i+\beta_i}{\alpha_r+\beta_r}\gamma_r\Delta\mathcal{F}^{\parallel}-\xi^{\parallel}_i\mathcal{G}^2.
\end{align}
\end{subequations}

In the transverse case, \eqref{beating equation} reduces to 
\begin{equation}\label{beating equation trans}
(\alpha+\beta)B^2+\gamma\Delta\mathcal{F}
-\xi^{\perp}\mathcal{G}^2-i\nu=0,
\end{equation}
which is the same as \eqref{beating equation long} except that $\xi^{\parallel}$ is replaced by $\xi^{\perp}$. Hence, using (\ref{thresholds}b), there are solutions for $\Delta\mathcal{F}>\Delta\mathcal{F}^{\perp}$, denoted  $B=B^{\perp}$ and $\nu=\nu^{\perp}$, given by 
\begin{subequations}
\label{mag nu trans}
\begin{align}
B^{\perp}&=\sqrt{\frac{\gamma_r(\Delta\mathcal{F}-\Delta\mathcal{F}^\perp)}{-(\alpha_r+\beta_r)}}, \\ 
\nu^{\perp}&=\left(\gamma_i-\frac{\alpha_i+\beta_i}{\alpha_r+\beta_r}\gamma_r\right)\Delta\mathcal{F}+\frac{\alpha_i+\beta_i}{\alpha_r+\beta_r}\gamma_r\Delta\mathcal{F}^{\perp}-\xi^{\perp}_i\mathcal{G}^2.
\end{align}
\end{subequations}

The longitudinal- and transverse-beating states exist for $\mathcal{F}$ above their respective thresholds \eqref{thresholds}; they are parameterized by a real-valued phase $\vartheta$ [cf.~\eqref{beating form}]. In the far-from-onset (or very-weak-shear) limit $\Delta\mathcal{F}/\mathcal{G}^2\to\infty$, these beating states approach the shear-free beating states \eqref{beating free} in magnitude and frequency, though not necessarily in direction: for arbitrarily weak shear, \emph{steady} beating is only possible in the longitudinal or transverse directions, whereas in the absence of shear any horizontal direction is permissible.  

\subsection{Stability of the longitudinal- and transverse-beating states}
\label{ssec:beatingstability}

\subsubsection{Longitudinal beating is unstable}
Consider the stability of the longitudinal-beating states, which exist for $\Delta\mathcal{F}>\Delta\mathcal{F}^{\parallel}$ and correspond to quasi-steady solutions \eqref{quasi} with $\bar{\bA}=\be_xB^{\parallel}\exp(i\vartheta)$ and $\nu=\nu^{\parallel}$, where $B^{\parallel}$ and $\nu^{\parallel}$ are provided by \eqref{mag nu long} and $\vartheta$ is a phase parameter. Upon substituting this solution form into the linearized amplitude equation \eqref{linearised about quasisteady}---setting $\vartheta=0$ without loss of generality---we find that the Cartesian components of the perturbation satisfy a pair of \emph{decoupled} differential equations, 
\begin{subequations}
\label{ax ay long}
\begin{align}
\frac{da_x}{dt}&=(\alpha+\beta){B^{\parallel}}^2(a_x+a_x^*),\\
\frac{da_y}{dt}&=[(\xi^{\parallel}-\xi^{\perp})\mathcal{G}^2-\alpha{B^{\parallel}}^2]a_y+\alpha{B^{\parallel}}^2a_y^*.
\end{align}
\end{subequations}
These respectively govern the growth of longitudinal perturbations (deformations parallel to the beating plane) and transverse perturbations (deformations normal to the beating plane). 

\medskip
\noindent\textbf{Longitudinal (in-plane) perturbations.} Since $a_x+a_x^*=2a_{x,r}$, (\ref{ax ay long}a) gives the pair of differential equations
\refstepcounter{equation} 
$$
\label{ax sys long}
\frac{da_{x,r}}{dt}=2(\alpha_r+\beta_r){B^{\parallel}}^2a_{x,r}, \quad 
\frac{da_{x,i}}{dt}=2(\alpha_i+\beta_i){B^{\parallel}}^2a_{x,r}. 
\eqno{(\theequation \mathrm{a},\mathrm{b})}
$$
The general solution of this system is 
\begin{equation}\label{ax sys long sol}
\begin{pmatrix}
a_{x,r}\\a_{x,i}
\end{pmatrix}
=c_1e^{2(\alpha_r+\beta_r){B^{\parallel}}^2t}
\begin{pmatrix}
\alpha_r+\beta_r\\ \alpha_i+\beta_i
\end{pmatrix}
+c_2
\begin{pmatrix}
0\\1
\end{pmatrix}
,
\end{equation}
where $c_1$ and $c_2$ are arbitrary real constants. The growth rates are therefore 
\refstepcounter{equation}
\label{ax sigma long}
$$
\sigma\ub{1}=0, \qquad \sigma\ub{2}=2(\alpha_r+\beta_r){B^{\parallel}}^2.
\eqno{(\theequation \mathrm{a},\mathrm{b})}
$$
Since $\alpha_r$ and $\beta_r$ are negative, the nonvanishing growth rate $\sigma\ub{2}$ is negative. Hence, the longitudinal-beating states are stable under longitudinal (in-plane) perturbations up to a neutral mode representing infinitesimal phase shifts. 

\medskip
\noindent\textbf{Transverse (out-of-plane) perturbations.}
Similarly, (\ref{ax ay long}b) yields the  system
\begin{equation}
\frac{d}{dt}\begin{pmatrix}a_{y,r} \\a_{y,i}\end{pmatrix}
=
\begin{pmatrix}
\kappa_r & 2{B^{\parallel}}^2\alpha_i-\kappa_i \\
\kappa_i & -2{B^{\parallel}}^2\alpha_r+\kappa_r\end{pmatrix}\begin{pmatrix}a_{y,r} \\a_{y,i}\end{pmatrix},
\end{equation}
where for brevity we define 
$\kappa=(\xi^{\parallel}-\xi^{\perp})\mathcal{G}^2$. Seeking exponential solutions [cf.~\eqref{exp ansatz}] yields the dispersion relation
\begin{equation}
\label{quadratic ay long}
\sigma^2+2(\alpha_r{B^{\parallel}}^2-\kappa_r)\sigma + \kappa_r^2+\kappa_i^2-2{B^{\parallel}}^2(\alpha_r\kappa_r+\alpha_i\kappa_i)=0.
\end{equation}
This is a quadratic equation for the $\sigma$ growth rates,  with real coefficients. Since $\alpha_r<0$ and $\kappa_r>0$, the coefficient $2({B^{\parallel}}^2\alpha_r-\kappa_r)$ is  negative. From Vieta's formulas, the sum of the roots is thus positive, implying that at least one root necessarily lies in the right half-plane. It follows that the longitudinal beating states are unstable under transverse (out-of-plane) perturbations. We conclude that the longitudinal-beating states are unstable, though if the deformation is restricted to the $xz$ plane then these states are stable up to a neutral mode corresponding to phase shifts.

\subsubsection{Transverse beating is stable up to a secondary bifurcation }
We next analyze the stability of the transverse-beating states, which exist for $\Delta\mathcal{F}>\Delta\mathcal{F}^{\perp}$ and correspond to quasi-steady solutions \eqref{quasi} with $\bar{\bA}=\be_yB^{\perp}\exp(i\vartheta)$ and $\nu=\nu^{\perp}$, where $B^{\perp}$ and $\nu^{\perp}$ are provided by \eqref{mag nu trans} and $\vartheta$ is a phase parameter. Upon substituting this solution form into the linearized amplitude equation \eqref{linearised about quasisteady}---setting $\vartheta=0$ without loss of generality---we here too find that the Cartesian components of the perturbation satisfy a pair of \emph{decoupled} differential equations,
\begin{subequations}
\label{ax ay trans}
\begin{align}
\frac{da_x}{dt}&=[(\xi^{\perp}-\xi^{\parallel})\mathcal{G}^2-\alpha{B^{\perp}}^2]a_x+\alpha{B^{\perp}}^2a_x^*,\\ 
\frac{da_y}{dt}&=(\alpha+\beta){B^{\perp}}^2(a_y+a_y^*),
\end{align}
\end{subequations}
respectively governing longitudinal perturbations (now representing deformations orthogonal to the beating plane) and transverse perturbations (now representing deformations parallel to the beating plane). 

\medskip
\noindent\textbf{Transverse (in-plane) perturbations.}
Equation (\ref{ax ay trans}b) governing $a_y(t)$ is of the same form as (\ref{ax ay long}a), the equation governing $a_x(t)$ in the longitudinal-beating analysis, except that $B^\parallel$ is replaced by $B^{\perp}$. Accordingly, the $\sigma$ growth rates in the present case immediately follow from \eqref{ax sigma long},  
\refstepcounter{equation}
$$
\sigma\ub{1}=0, \qquad \sigma\ub{2}=2{B^\perp}^2(\alpha_r+\beta_r).
\eqno{(\theequation \mathrm{a},\mathrm{b})}
$$
Hence, the transverse-beating states are stable under transverse (in-plane) perturbations, up to phase shifts.

\medskip
\noindent\textbf{Longitudinal (out-of-plane) perturbations.} Equation (\ref{ax ay trans}a) governing $a_x(t)$ is of the same form as (\ref{ax ay long}b), the equation governing $a_y(t)$ in the longitudinal-beating analysis, except that $B^\parallel$ and $(\xi^{\parallel}-\xi^{\perp})$ are replaced by $B^{\perp}$ and $(\xi^{\perp}-\xi^{\parallel})$, respectively. Hence, the $\sigma$ growth rates in the present case are governed by the dispersion relation [cf.~\eqref{quadratic ay long}]
\begin{equation}
\label{quadratic ax trans}
\sigma^2+2(\alpha_r{B^{\perp}}^2+\kappa_r)\sigma + \kappa_r^2+\kappa_i^2+2{B^{\perp}}^2(\alpha_r\kappa_r+\alpha_i\kappa_i)=0.
\end{equation}

\begin{figure}[t]
\begin{center}
\includegraphics[scale=0.5]{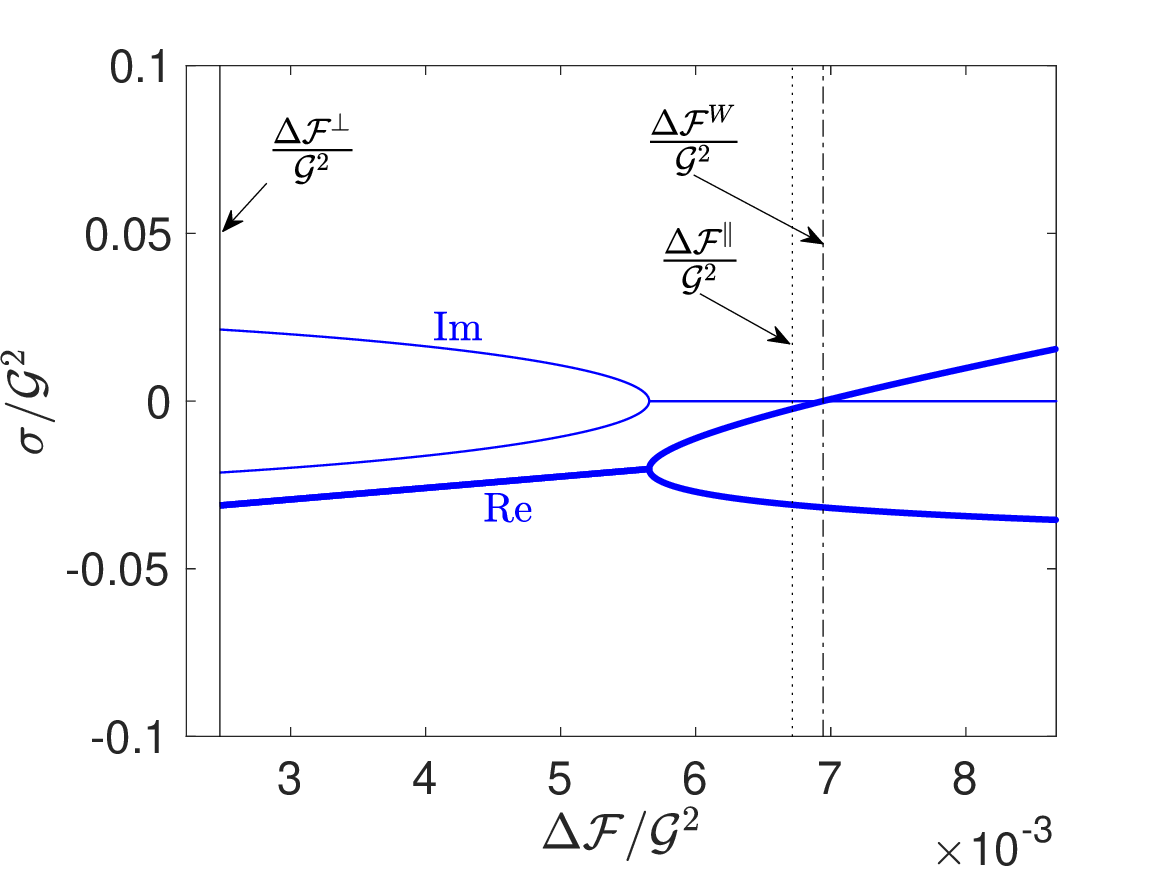}
\caption{Growth rates for transverse-beating states under longitudinal (out-of-plane) perturbations.}
\label{fig:stabTransXpert}
\end{center}
\end{figure}

The argument used below \eqref{quadratic ay long} to prove instability does not generally apply here, as the sign of $\alpha_r{B^{\perp}}^2+\kappa_r$ depends on the parameters $\Delta\mathcal{F}$ and $\mathcal{G}$ (through $B^{\perp}$ and $\kappa$). In fact, the $\sigma$ growth rates, when  normalized by $\mathcal{G}^2$, are functions solely of the control parameter $\Delta\mathcal{F}/\mathcal{G}^2$. In Fig.~\ref{fig:stabTransXpert}, we plot these functions for the two roots of \eqref{quadratic ax trans}. Upon increasing $\Delta\mathcal{F}/\mathcal{G}^2$ from zero, the roots initially form a complex-conjugate pair with a negative real part. They subsequently transform into a negative-real pair, with one root increasing and the other decreasing. At a critical threshold, the larger root becomes positive, meaning that the transverse-beating states become unstable under longitudinal (out-of-plane) perturbations. Thus, beyond this threshold, the transverse-beating states are generally unstable, though they remain neutrally stable under transverse (in-plane) perturbations. 

\medskip
\noindent\textbf{Secondary bifurcation.}
The fact that the transverse-beating states destabilize via a real growth rate crossing the origin---rather than a complex-conjugate pair crossing the imaginary axis---has two important consequences. First, it means that we can readily obtain the instability threshold by setting $\sigma=0$ in \eqref{quadratic ax trans}. This gives the corresponding critical value of $\Delta\mathcal{F}$ as
\begin{equation}\label{threshold W}
\Delta\mathcal{F}^{W}=\Delta\mathcal{F}^{\perp}+\mathcal{G}^2\frac{\alpha_r+\beta_r}{2\gamma_r}\frac{|\xi^{\parallel}-\xi^{\perp}|^2}{\mathrm{Re}[\alpha^*(\xi^{\parallel}-\xi^{\perp})]},
\end{equation}
where we have used (\ref{thresholds}b) and (\ref{mag nu trans}a). This threshold happens to be numerically close to that for instability of the steady-deflection state under longitudinal perturbations: $(\Delta\mathcal{F}^W-\Delta\mathcal{F}^{\parallel})/\mathcal{G}^2\approx 0.0069$  [cf.~\eqref{thresholds}]. Heuristically, this has to do with the fact that both of these thresholds are associated with longitudinal beating becoming unstable. 

Second, the neutral longitudinal perturbations at the threshold \eqref{threshold W} oscillate with the same angular frequency as the transverse-beating states, $\omega+\nu^{\perp}$. From \eqref{perturbation from beating}, these neutral perturbations can be written as 
\begin{equation}\label{elliptical threshold}
\bA(t)-e^{i\nu^{\perp}t}e^{i\vartheta}B^{\perp}\be_y\approx \mathcal{W} e^{i\nu^{\perp}t}\be_x,
\end{equation}
where $\mathcal{W}$ is an arbitrary (infinitesimally small) complex constant. Thus, the superposition of the emergent longitudinal mode onto the   transverse-beating base state produces linear elliptical whirling with near-unity eccentricity and near-transverse orientation of the major axis.

Given the symmetry about the $xz$ plane, the above observations suggest that clockwise and counterclockwise elliptical-whirling states emerge from the transverse-beating states at the secondary threshold $\Delta\mathcal{F}=\Delta\mathcal{F}^W$ via a pitchfork bifurcation of cycles. In principle, this could be demonstrated analytically via a nested weakly nonlinear analysis, of the amplitude equation \eqref{amplitude shear} near this bifurcation; such an analysis would show that \eqref{elliptical threshold} holds near the elliptical-whirling threshold, with the complex constant $\mathcal{W}$ evolving on a super-slow time scale according to a reduced amplitude equation. Analysis of this amplitude equation, in turn, would provide a near-onset approximation for the elliptical-whirling states and reveal their stability in that regime. We do not pursue this local analysis herein. Rather, in the next subsection, we calculate the elliptical-whirling states by numerically solving the steady amplitude equation \eqref{amplitude steady} and subsequently determine their stability based on an analysis of the linearized amplitude equation \eqref{linearised about quasisteady}.
\begin{figure}[t!]
\begin{center}
\includegraphics[scale=0.58,trim={1.3cm 0.5cm 0 0}]{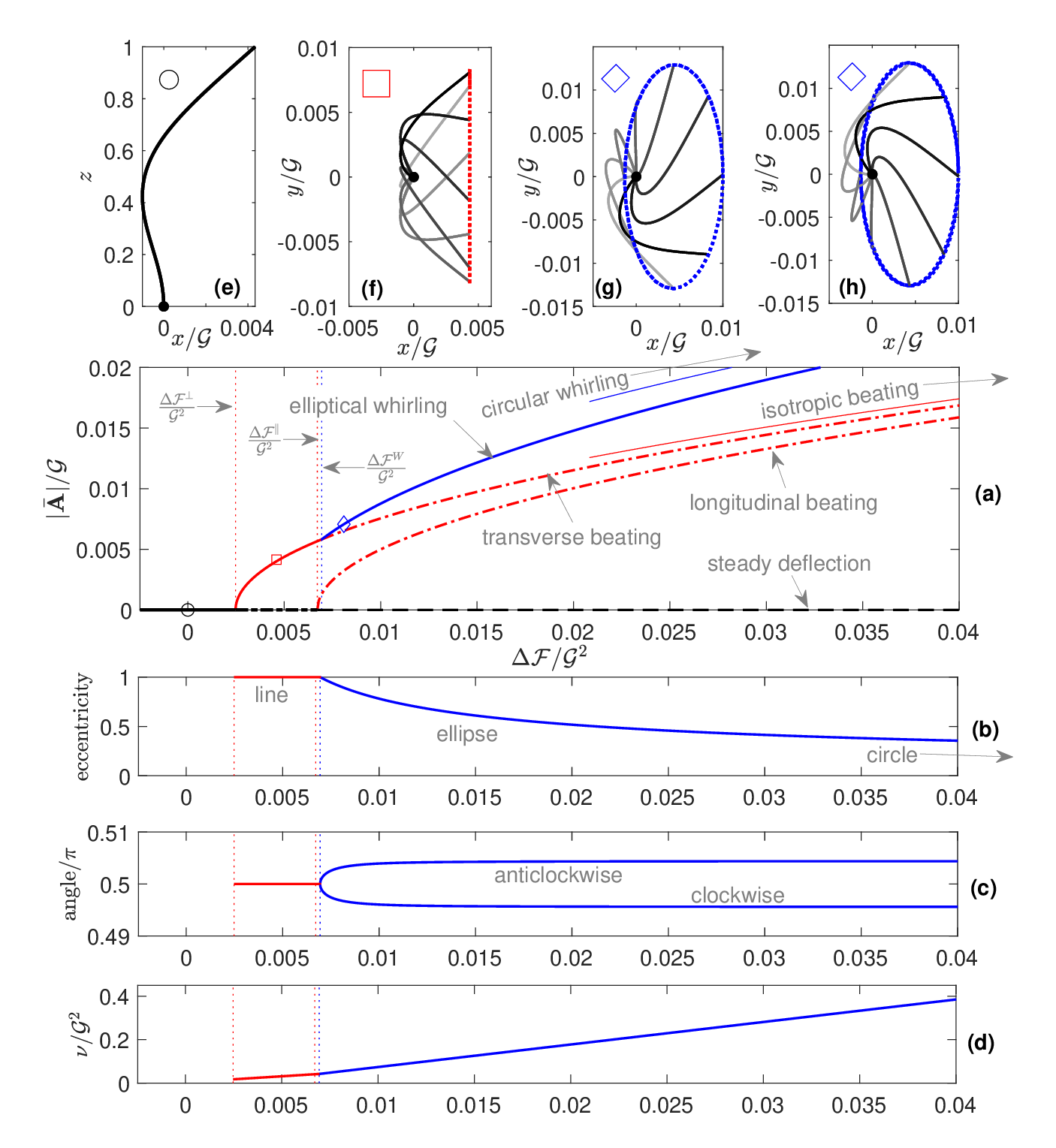}
\caption{(a) Bifurcation diagram predicted by the weakly nonlinear theory. Solid: stable or neutrally stable states. Dash-dotted: unstable, but neutrally stable under parallel perturbations. Dashed: unstable. (b) Eccentricity of the tip trajectory for the transverse-beating and elliptical-whirling states. (c) Angle from the longitudinal direction to the major axis of the tip ellipse. (d) Angular-frequency correction for the transverse-beating and elliptical-whirling states. (e-h) Depiction of the steady-deflection, transverse-beating and clockwise and counter-clockwise elliptical-whirling states marked in (a). In (f-h), the greyscale solid curves show horizontal projections of the filament centerline at equally spaced times over the period, and the dotted curves trace the tip orbits.}
\label{fig:sheareffects}
\end{center}
\end{figure}
\begin{figure}[t!]
\begin{center}
\includegraphics[scale=0.5]{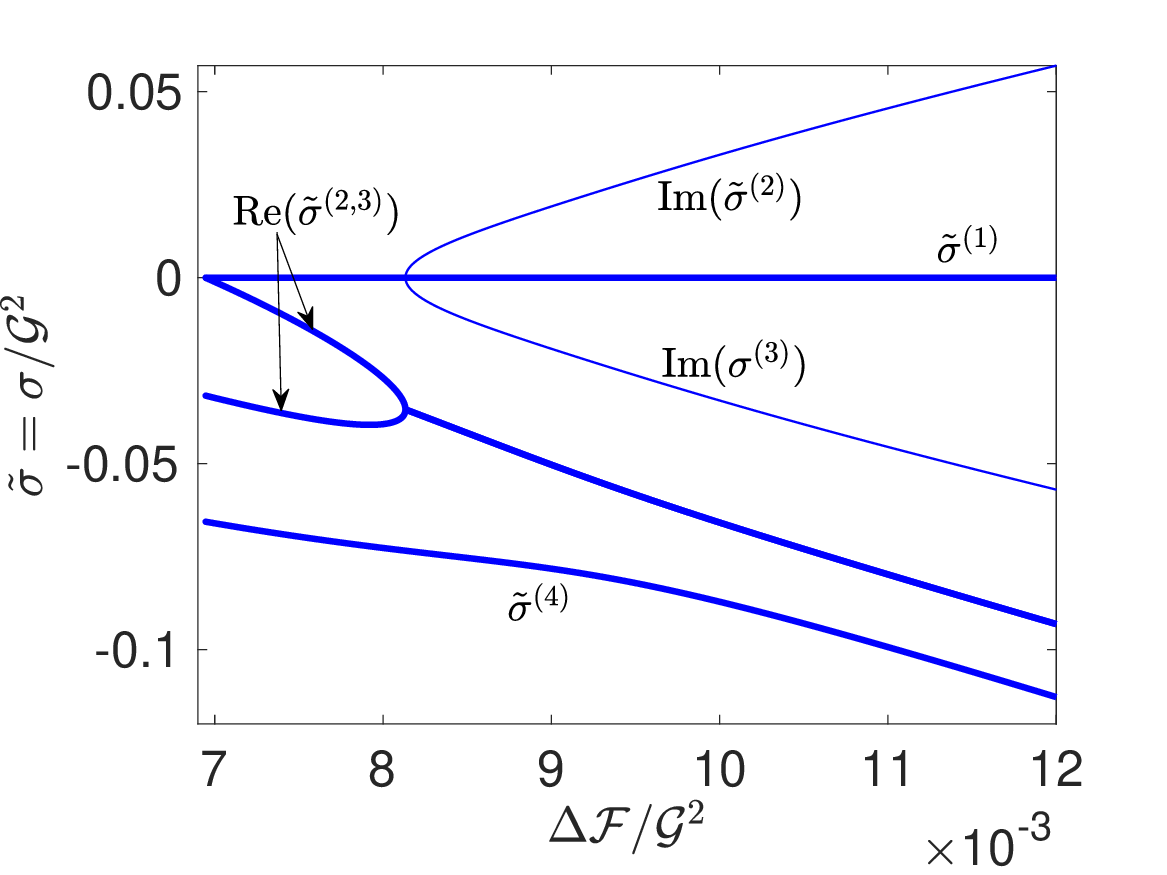}
\caption{Growth rates $\sigma$ of perturbations about the elliptical-whirling states. Besides the zero growth rate associated with neutrality under phase shifts, the growth rates have a negative real part for all values of the control parameter $\Delta\mathcal{F}/\mathcal{G}^2$ for which these states exist.}
\label{fig:EWstab}
\end{center}
\end{figure}

\subsection{Elliptical-whirling states and overall dynamics}
\label{ssec:ewhirling}
We search for the elliptical-whirling states anticipated in the preceding subsection by numerically solving the steady amplitude equation \eqref{amplitude steady} written as a system of four nonlinear algebraic equations for the five unknowns: the time-independent components $\bar{A}_{x,r}$, $\bar{A}_{x,i}$, $\bar{A}_{y,r}$ and $\bar{A}_{y,i}$, and the angular-frequency correction $\nu$. The under-determinacy arises from the arbitrariness of the phase along the trajectory. In our calculation, we remove this arbitrariness by fixing one of the first four components. Since $\bar{\bA}/\mathcal{G}$ depends solely on the control parameter $\Delta\mathcal{F}/\mathcal{G}^2$, it suffices to solve for $\mathcal{G}=1$, varying $\Delta\mathcal{F}$. Anticipating that the elliptical-whirling states evolve, with increasing $\Delta\mathcal{F}$, into the circular-whirling states found in the shear-free case, we use the circular-whirling solution \eqref{whirling free} as an initial guess at large $\Delta\mathcal{F}$ and perform continuation down to the secondary bifurcation at $\Delta\mathcal{F}=\Delta\mathcal{F}^W$ from the transverse-beating states [cf.~\eqref{threshold W}]. Multiplying a computed solution $\bar{\bA}$ by a factor $\exp(i\vartheta)$, where $\vartheta$ is an arbitrary real constant, generates a family of states that differ only in phase. Furthermore, the direction of travel of the computed state is inherited from the initial circular-whirling guess; a counter-rotating state may be obtained by flipping the sign of $\bar{A}_y$, giving an elliptical orbit reflected about the $x$-axis. Last, the linear stability of computed states is determined following the general approach described in \S\ref{ssec:amplitude}, see \eqref{perturbation from beating} \textit{et seq.}; this leads to a $4\times4$ matrix eigenvalue problem which is readily solved numerically. 

The bifurcation diagram we find is shown in  Fig.~\ref{fig:sheareffects}, alongside an example steady-deflection profile, example transverse-beating and elliptical-whirling motions, and plots of the eccentricity, orientation and frequency of the elliptical-whirling trajectories. We scale amplitudes and horizontal displacements by $\mathcal{G}$ and frequency by $\mathcal{G}^2$, so that the results  depend solely on the control parameter $\Delta\mathcal{F}/\mathcal{G}^2$. As anticipated, there is a continuous branch of elliptical-whirling states existing for $\Delta\mathcal{F}>\Delta\mathcal{F}^W$. The eccentricity varies from unity at $\Delta\mathcal{F}^W$  (transverse beating) to zero as $\Delta\mathcal{F}\to\infty$ (circular whirling). During this transition, the elliptical trajectories rotate, in opposite directions for the clockwise and counter-clockwise states, whose orbits differ from one another, unlike in the  circular-whirling case. The elliptical-whirling states are always stable up to neutrality under phase shifts; see Fig.~\ref{fig:EWstab} for a plot of the perturbation growth rates. 

The bifurcation diagram implies that, as $\Delta\mathcal{F}/\mathcal{G}^2$ is decreased down to zero---say, by increasing the shear rate at a fixed follower force above the shear-free instability threshold---we generally expect a transition from circular to elliptical whirling, then to transverse beating, and ultimately to steady deflection. If the initial conditions restrict the deformation to the $xz$-plane, however, we expect a transition from longitudinal beating to steady deflection. We confirmed these dynamics via initial-value simulations of the amplitude equation \eqref{amplitude shear}, which we omit as they provide no additional insight beyond what the bifurcation diagram conveys.

\section{Concluding remarks}\label{sec:conclusions}

Our analysis demonstrates that, in the inertialess Stokes-flow limit and within the weakly nonlinear near-critical and weak-shear regime, an externally imposed simple shear flow effectively damps the spontaneous oscillations of a follower-force active filament clamped normally to a wall. This damping mechanism arises from a subtle nonlinear resonance between the imposed steady flow and the intrinsic whirling modes of the filament. Owing to the nature of this resonance, the damping is linear in the oscillation amplitude, with  proportionality coefficients depending quadratically on the shear rate. Furthermore, this damping is anisotropic, with the resistance to  longitudinal oscillations being nearly three times stronger than that to transverse oscillations. 

The physical effect of this anisotropic damping is most clearly illustrated by considering a scenario where the shear rate is gradually increased from zero at a fixed follower-force magnitude just beyond the shear-free instability threshold. In the absence of shear, circular whirling is expected (for most initial conditions). At very small shear rates, the anisotropy does not fundamentally suppress the motion but rather deforms the circular whirling into elliptical whirling, with the major axis of the tip trajectory oriented nearly perpendicular to the shear flow. As the shear rate is increased, the fact that longitudinal motion is damped more strongly than transverse motion renders the ellipse increasingly eccentric. At a critical shear threshold, the elliptical tip orbit collapses onto a line segment normal to the shear flow, corresponding to a planar transverse-beating state. Beyond a second critical threshold, even these transverse oscillations are completely damped, leaving the filament in a steady deflection state.

This shear-induced stabilization resembles a trend observed in considerably more complex models, of cytoplasmic streaming, where arrays of follower-force filaments interact within a confined geometry \citep{Stein:21,Dutta:24}. For sufficiently dilute arrays, filaments exhibit independent whirling oscillations analogous to the single-filament problem. However, beyond a critical density (dependent upon the follower-force magnitude), collective interactions trigger a symmetry-breaking transition into steady streaming. In that streaming regime, the filament oscillations are suppressed. As the streaming flow can be locally approximated by a simple shear flow \citep{Stein:21}, it seems plausible that this stabilization is governed by the same shear-induced damping mechanism illustrated by the present minimal model. If so, our theory suggests that an intermediate dynamical regime---wherein filaments oscillate transversely to the local macroscopic flow---may be observable in these cytoplasmic simulations. 

Reproducing this collective-streaming scenario would entail both generalizing our analysis to arrays of interacting filaments and adding a crucial element to the physical model. Indeed, if the follower force models the force exerted by a translocating molecular motor on the filament, then there is also a reaction force on the motor; since the motor is itself inertialess, this implies a reverse follower force acting on the liquid. (This entrainment force is accounted for in the filament-array simulations \citep{Stein:21,Dutta:24} and in the single-filament planar analysis of \citet{De:17}.) Within a weakly nonlinear framework, filament interactions could introduce two distinct resonance mechanisms: a linear resonance with the oscillatory flow generated by the periodic motion of neighbouring filaments, and a nonlinear resonance with the emergent steady streaming flow. As a reduced model for the former, it would be instructive to analyze a single filament subjected to an imposed \emph{oscillatory} shear flow with a frequency near the natural instability frequency, where linear resonance is expected to produce interactions qualitatively stronger than the nonlinear resonance found here under steady shear. 

Finally, while our analysis focused on the near-onset, weak-shear regime, perturbation methods could also be used to address weak-shear effects \emph{away} from onset. Rather than a weakly nonlinear analysis, this regime would involve expanding about numerically calculated beating or whirling states to capture the slow phase evolution along these limit cycles or the reorientation of the beating plane. Similarly, near-onset dynamics under moderate shear could be studied by expanding about a numerically calculated fully nonlinear steady-deflection state, with the instability threshold significantly shifted from its shear-free value. More generally, adapting the computational framework of \citet{Clarke:24} would allow exploration of arbitrary parameter regimes and incorporation of nonlocal hydrodynamic effects. Conversely, the asymptotic methods employed in the present paper could be used to analyze the preferred-curvature and cross-sectional anisotropy effects recently studied numerically by that group \citep{Clarke:25}.

\backmatter

%\bmhead{Acknowledgements}

%Acknowledgements are not compulsory. Where included they should be brief. Grant or contribution numbers may be acknowledged.

%\section*{Declarations}

%\begin{itemize}
%\item Funding `Not applicable' 
%\item Conflict of interest/Competing interests `Not applicable' 
%\item Ethics approval and consent to participate `Not applicable' 
%\item Consent for publication `Not applicable' 
%\item Data availability `Not applicable'  
%\item Materials availability `Not applicable' 
%\item Code availability `Not applicable'  
%\item Author contribution `Not applicable' 
%\end{itemize}

%\noindent
%If any of the sections are not relevant to your manuscript, please include the heading and write `Not applicable' for that section. 


\begin{appendices}
\section{Derivation of the amplitude equation}\label{app:WNA}
This appendix presents a derivation of the amplitude equation \eqref{amplitude shear} by extending the weakly nonlinear analysis in \citep{Schnitzer:25filament} of the shear-free case and building on the scaling arguments in \S\ref{ssec:separation}. The derivation is so technically dreadful that it may seem unlikely that it is free of calculation errors. Thankfully, using a code written with the help of Gemini 3.1 Pro \citep{Gemini} (Supplementary Information, Part II), we are able to reproduce the entire weakly nonlinear analysis in Mathematica \citep{Mathematica}.

\subsection{Weakly nonlinear expansions}
Following \citep{Schnitzer:25filament}, we write the dimensionless follower force as 
\begin{equation}\label{chi def}
\mathcal{F}=\mathcal{F}_c+\epsilon \chi,
\end{equation}
where $\mathcal{F}_c$ is the critical value of the follower force in the absence of shear [cf.~\S\ref{ssec:homogeneous}], $\epsilon$ is a positive small parameter and $\chi$ is a rescaled bifurcation parameter. Following the scaling arguments in \S\ref{ssec:separation}, we scale the shear rate as
\begin{equation}
\label{g def}
\mathcal{G}=\epsilon^{1/2}g,
\end{equation} 
with $g$ introduced as a reduced shear rate. In the main text, the product $\epsilon\chi$ is denoted by $\Delta\mathcal{F}$. It is only this product, rather than $\epsilon$ or $\chi$ separately, that is physically meaningful. For $\mathcal{F}\ne\mathcal{F}_c$, we could scale out $|\chi|$ by choosing $\epsilon=|\Delta\mathcal{F}|$, reducing the parameter set to $\chi$ restricted to $\pm1$ and $g>0$. Alternatively, for $\mathcal{G}>0$, we could scale out $g$ by choosing $\epsilon=\mathcal{G}^2$, leaving $\chi=\Delta\mathcal{F}/\mathcal{G}^2$ as the sole control parameter. We find it convenient to retain both $\chi$ and $g$ as real-valued parameters in the derivation below, to help identify terms associated with the imposed shear flow and to simplify the consideration of the special cases where either $\Delta\mathcal{F}=0$ or $\mathcal{G}=0$. 

With these definitions, the distinguished near-onset and weak-shear limit identified in \S\ref{ssec:separation} corresponds to $\epsilon\to0$ with $\chi,g$ held fixed, oscillation amplitudes scaling as $\epsilon^{1/2}$ and a slow time scale scaling as $1/\epsilon$. As in \citep{Schnitzer:25filament}, we introduce the two-scale extension $\underline{\br}(s,\tau,T)$ of the centreline $\br(s,t)$, where the ``fast time'' $\tau$ and ``slow time'' $T$ are treated as independent coordinates such that $\underline{\br}(s,\tau,T)=\br(s,t)$ on the ``physical diagonal'': 
\begin{equation}\label{diagonal}
\tau=t \quad \text{and} \quad T=\epsilon t.
\end{equation}
Analogous extensions are introduced for the remaining fields, denoted $\underline{\bt}(s,\tau,T)$, $\underline{\bF}(s,\tau,T)$ and $\underline{\bM}(s,\tau,T)$. The extended fields satisfy the same problem formulated in \S\ref{sec:formulation}, with the time derivative in (\ref{force and moment balances}a) transformed as
\begin{equation}\label{time transform}
\pd{}{t}\Rightarrow \pd{}{\tau}+\epsilon\pd{}{T}.
\end{equation}

For ease of reference, we write the full problem for the extended fields derived from the governing equations \eqref{force and moment balances}--\eqref{bcs tip}, incorporating the rescalings \eqref{chi def} and \eqref{g def}, and the transformation \eqref{time transform}. It consists of the partial differential equations
\begin{subequations}
\label{pdes transformed}
\begin{gather}
\pd{\underline{\br}}{s}=\underline{\bt},\\
 \pd{\underline{\bF}}{s}-\left(\tI-\frac{1}{2}\underline{\bt}\underline{\bt}\right)\bcdot \pd{\underline{\br}}{\tau}-\epsilon\left(\tI-\frac{1}{2}\underline{\bt}\underline{\bt}\right)\bcdot \pd{\underline{\br}}{T}=-\epsilon^{1/2}\left(\tI-\frac{1}{2}\underline{\bt}\underline{\bt}\right)\bcdot g z\be_x\\
\pd{\underline{\bM}}{s}+\underline{\bt}\times\underline{\bF}=\bzero, \\  \underline{\bM}=\underline{\bt}\times\pd{\underline{\bt}}{s};
\end{gather}
\end{subequations}
the geometric constraint 
\begin{equation}
\label{constraint transformed}
\underline{\bt}\bcdot \underline{\bt}=1;
\end{equation}
the boundary conditions at the clamping point,
\refstepcounter{equation}
$$
\label{bcs wall transformed}
\underline{\br}=\bzero, \quad \underline{\bt}=\be_z \qquad \text{at} \quad s=0;\eqno{(\mathrm{A}\arabic{equation} \mathrm{a},\mathrm{b})}
$$
and the boundary conditions at the tip, 
\refstepcounter{equation}
$$
\label{bcs tip transformed} 
\underline{\bF}= -\mathcal{F}_c\underline{\bt}-\epsilon \chi\underline{\bt}, \qquad \underline{\bM}=\bzero \quad \text{at} \quad s=1.  \eqno{(\mathrm{A}\arabic{equation} \mathrm{a},\mathrm{b})}
$$

We next expand the extended vector fields about the vertical steady state evaluated at $\mathcal{F}=\mathcal{F}_c$,
\begin{subequations}
\label{wna expansions}
\begin{eqnarray}
\underline{\br}(s,\tau,T)&=&
s\be_z+\epsilon^{1/2}\br_{1/2}(s,\tau,T)+\epsilon \br_1(s,\tau,T)+\epsilon^{3/2}\br_{3/2}(s,\tau,T)+\cdots,\\
\underline{\bt}(s,\tau,T)&=&
\be_z+\epsilon^{1/2}\bbt_{1/2}(s,\tau,T)+\epsilon \bbt_1(s,\tau,T)+\epsilon^{3/2}\bbt_{3/2}(s,\tau,T)+\cdots,\\
\underline{\bF}(s,\tau,T)&=&
-\mathcal{F}_c\be_z+\epsilon^{1/2}\bF_{1/2}(s,\tau,T)+\epsilon \bF_1(s,\tau,T)+\epsilon^{3/2}\bF_{3/2}(s,\tau,T)+\cdots,\\
\underline{\bM}(s,\tau,T)&=&
\epsilon^{1/2}\bM_{1/2}(s,\tau,T)+\epsilon \bM_1(s,\tau,T)+\epsilon^{3/2}\bM_{3/2}(s,\tau,T)+\cdots.
\end{eqnarray}
\end{subequations}
We note that the progression in half-powers of $\epsilon$ is hinted at by the expected scaling of the oscillation, and that the terms $\bbt_{1/2}$, $\bbt_1$, $\bbt_{3/2}$,$\ldots$ are not necessarily unit vectors and are therefore not decorated by hats. Following the method of multiple scales \citep{Hinch:91}, the additional degrees of freedom introduced by treating $\tau$ and $T$ as independent variables are removed by requiring that \eqref{wna expansions} hold as regular asymptotic expansions as $\epsilon\to0$. 

From this point, the analysis closely follows that in \citep{Schnitzer:25filament}, with extra terms arising from the shear-induced drag on the right-hand side of (\ref{pdes transformed}b). To facilitate comparison with the shear-free derivation \citep{Schnitzer:25filament}, we note that this term possesses the expansion
\begin{multline}
\label{shear expansion}
-\epsilon^{1/2}\left(\tI-\frac{1}{2}\underline{\bt}\underline{\bt}\right)\bcdot gz\be_x = -\epsilon^{1/2}gs\be_x+\epsilon g\left[-z_{1/2}\be_x+\frac{1}{2}s(\be_x\bcdot\bbt_{1/2})\be_z\right]\\
+\epsilon^{3/2}g\left[-z_1\be_x+\frac{1}{2}z_{1/2}(\bbt_{1/2}\bcdot\be_x)\be_z+\frac{1}{2}s(\be_x\bcdot\bbt_1)\be_z+\frac{1}{2}s(\be_x\bcdot\bbt_{1/2})\bbt_{1/2}\right] + \cdots
\end{multline}
where $z_{1/2}=\be_z\bcdot\br_{1/2}$, $z_1=\be_z\bcdot\br_1$,$\ldots$ 

\subsection{The $\ord(\epsilon^{1/2})$ problem}
At $\ord(\epsilon^{1/2})$, the governing differential equations \eqref{pdes transformed} give
\begin{subequations}
\label{eqs 1/2}
\begin{align}
\pd{\br_{1/2}}{s}-\bbt_{1/2}&=\bzero,  \\
 \pd{\bF_{1/2}}{s}-\left(\tI-\frac{1}{2}\be_z\be_z\right)\bcdot \pd{\br_{1/2}}{\tau}&=-gs\be_x, \\ 
\pd{\bM_{1/2}}{s}+\mathcal{F}_c\be_z\times\bbt_{1/2}+\be_z\times\bF_{1/2}&=\bzero, \\ 
\bM_{1/2}-\be_z\times\pd{\bbt_{1/2}}{s}&=\bzero;
\end{align}
\end{subequations}
the constraint \eqref{constraint transformed} gives 
\begin{equation}\label{constraint 1/2}
\be_z\bcdot\bt_{1/2}=0;
\end{equation}
and the boundary conditions \eqref{bcs wall transformed} and \eqref{bcs tip transformed} give 
\begin{subequations}
\label{bcs 1/2}
\begin{gather*}
\br_{1/2}=\bzero, \quad \bbt_{1/2}=\bzero \quad \text{at} \quad s=0; \tag{\theequation a,b}\\
\bF_{1/2}+\mathcal{F}_c\bbt_{1/2}=\bzero, \quad \bM_{1/2}=\bzero \quad \text{at} \quad s=1. \tag{\theequation c,d}
\end{gather*}
\end{subequations}

The problem at this order represents the linearization of the exact problem about the vertical steady state under weak shear and at $\mathcal{F}=\mathcal{F}_c$, with standard time $t$ replaced by the fast time $\tau$. As such, this problem effectively corresponds to the linear theory in \S\ref{sec:linear}, considered at the threshold. Furthermore, it is identical to the $\ord(\epsilon^{1/2})$ problem in \citep{Schnitzer:25filament} except for the inhomogeneous (shear-induced) term in (\ref{eqs 1/2}b). The solution may therefore be written by inspection, drawing on the results summarized in \S\ref{sec:linear}. 

Since our interest is in the slow-time evolution of the dynamics, we discard linear modes that decay exponentially in the fast time $\tau$ (i.e., those with a negative real growth rate). The general solution is then given by the superposition (similarly for $\bbt_{1/2}$, $\bF_{1/2}$ and $\bM_{1/2}$),
\begin{equation}
\br_{1/2}(s,\tau,T)=\br_{1/2}^W(s,\tau,T)+\br_{1/2}^G(s),
\end{equation}
where $\br_{1/2}^W(s,\tau,T)$ is a solution of the homogeneous problem corresponding to elliptical whirling (but with an amplitude varying with the slow time $T$) and $\br_{1/2}^G(s)$ is a particular solution corresponding to steady deflection. 

Thus, from the results of \citep{Schnitzer:25filament} (also summarized in \S\ref{sec:linear}), we have 
\begin{subequations}
\label{sol 1/2}
\begin{align}
\br_{1/2}^W &= \bW(T)\varphi(s)e^{i\omega \tau}+\text{c.c.}, \\ 
\bbt_{1/2}^W &= \bW(T)\varphi'(s)e^{i\omega \tau}+\text{c.c.},\\
\bF_{1/2}^W &= -\bW(T)[\varphi'''(s)+\mathcal{F}_c\varphi'(s)]e^{i\omega \tau}+\text{c.c.}, \\ 
\bM_{1/2}^W &= \be_z\times\bW(T)\varphi''(s)e^{i\omega \tau}+\text{c.c.},
\end{align}
\end{subequations}
where $\bW(T)$ is a horizontal complex-vector amplitude and $\varphi(s)$ is an eigenfunction satisfying \eqref{phi def text}. (The amplitude $\bA$ in the main text corresponds to $\epsilon^{1/2}\bW$; in \citep{Schnitzer:25filament}, the amplitude $\tilde{\bA}$ corresponds to $\epsilon^{1/2}\bW$, whereas the amplitude $\bA$ corresponds to $\bW$.)

Furthermore, the particular solution $\br_{1/2}^G(s)$ follows from the results of \S\ref{sec:linear}; specifically, it is the steady-deflection solution \eqref{steady deflection} evaluated at $\mathcal{F}=\mathcal{F}_c$ and with $\mathcal{G}$ replaced by $g$: 
\begin{subequations}
\label{sol 1/2 G}
\begin{align}
\br_{1/2}^G&=g \ell_c(s)\be_x, \\ \bbt_{1/2}^G&=g\ell_c'(s)\be_x, \\ \bF_{1/2}^G&=\frac{g}{2}\left[1-s^2-2\mathcal{F}_c\ell_c'(1)\right]\be_x, \\ \bM_{1/2}^G&=g\ell_c''(s)\be_y,
\end{align}
\end{subequations}
where the profile $\ell_c(s)$ solves the boundary-value problem
\begin{equation}\label{lc problem}
\ell_c''''+\mathcal{F}_c\ell_c''=s, \quad \ell_c(0)=\ell_c'(0)=\ell_c''(1)=\ell_c'''(1)=0,
\end{equation} 
and is given explicitly by [cf.~\eqref{ell solution}] 
\begin{multline}\label{ell solution app}
\ell_c(s)=\frac{1}{6\mathcal{F}_c}s^3-\frac{\cos \sqrt{\mathcal{F}_c}}{\mathcal{F}_c^2}(1+s)+\frac{1}{\mathcal{F}_c^2}\cos[\sqrt{\mathcal{F}_c}(1-s)] \\
+\frac{\sin\sqrt{\mathcal{F}_c}}{\mathcal{F}_c^{5/2}}(1-\mathcal{F}_c s)-\frac{1}{\mathcal{F}_c^{5/2}}\sin[\sqrt{\mathcal{F}_c}(1-s)].
\end{multline}
The following relation will be useful at the next order:
\begin{equation}\label{ellc relation}
\ell_c'''+\mathcal{F}_c\ell_c'=\frac{1}{2}(s^2-1)+\mathcal{F}_c\ell_c'(1). 
\end{equation}

We note that the entire solution at $\ord(\epsilon^{1/2})$ is horizontal; thus $z_{1/2}=0$ in \eqref{shear expansion}.

\subsection{The $\ord(\epsilon)$ problem}
\subsubsection{Problem formulation}
At $\ord(\epsilon)$, the governing differential equations \eqref{pdes transformed} give, with the help of the shear-term expansion \eqref{shear expansion}, 
\begin{subequations}
\label{eqs 1}
\begin{eqnarray}
\pd{\br_{1}}{s}-\bbt_{1}&=&\bzero,  \\
 \pd{\bF_{1}}{s}-\left(\tI-\frac{1}{2}\be_z\be_z\right)\bcdot \pd{\br_{1}}{\tau}&=&\boldsymbol{\mathcal{R}}\ub{2}, \\ 
\pd{\bM_{1}}{s}+\mathcal{F}_c\be_z\times\bbt_{1}+\be_z\times\bF_{1}&=&\boldsymbol{\mathcal{R}}\ub{3}, \\ 
\bM_{1}-\be_z\times\pd{\bbt_{1}}{s}&=&\boldsymbol{\mathcal{R}}\ub{4}.
\end{eqnarray}
\end{subequations}
The forcing terms are 
\begin{subequations}
\begin{align}
\boldsymbol{\mathcal{R}}\ub{2} &= -\frac{1}{2}\left(\bbt_{1/2}\be_z+\be_z\bbt_{1/2}\right)\bcdot\pd{\br_{1/2}}{\tau}+\frac{g}{2}s(\be_x\bcdot\bbt_{1/2})\be_z,\\
\boldsymbol{\mathcal{R}}\ub{3} &= -\bbt_{1/2}\times\mathbf{F}_{1/2},\\
\boldsymbol{\mathcal{R}}\ub{4}&=\bbt_{1/2}\times\pd{\bbt_{1/2}}{s}. 
\end{align}
\end{subequations}
It is convenient to decompose these forcing terms as 
\begin{subequations}
\label{R2 decompositions}
\begin{align}
\boldsymbol{\mathcal{R}}\ub{2}&=\boldsymbol{\mathcal{R}}\ub{2,WW}+\boldsymbol{\mathcal{R}}\ub{2,GW}+\boldsymbol{\mathcal{R}}\ub{2,GG},\\
\boldsymbol{\mathcal{R}}\ub{3} &= \boldsymbol{\mathcal{R}}\ub{3,WW}+\boldsymbol{\mathcal{R}}\ub{3,GW},\\
\boldsymbol{\mathcal{R}}\ub{4}&=\boldsymbol{\mathcal{R}}\ub{4,WW}+\boldsymbol{\mathcal{R}}\ub{4,GW},
\end{align}
\end{subequations}
where the superscripts indicate the origin of the nonlinear forcing terms: $WW$ terms arise from the self-interaction of the leading-order whirling mode, $GW$ from the cross-interaction between the steady shear-induced deflection and the whirling, and $GG$ from the self-interaction of the steady shear-induced deflection. Analogous extensions of this notation will be employed at the next order.

By substituting the solutions from the preceding order, the different terms in (\ref{R2 decompositions}) can be written \begin{subequations}
\begin{align}
\boldsymbol{\mathcal{R}}\ub{2,WW}&=-\frac{1}{2}\be_z\bbt^W_{1/2}\bcdot\pd{\br_{1/2}^W}{\tau}
\notag \\ &=-\be_z\frac{i\omega}{2}\left(\bW\bcdot\bW\varphi'\varphi e^{2i\omega\tau}-|\bW|^2\varphi'\varphi^*\right)+\mathrm{c.c.},
\\
\boldsymbol{\mathcal{R}}\ub{2,GW}&=-\frac{1}{2}\be_z\bbt^G_{1/2}\bcdot\pd{\br_{1/2}^W}{\tau}
+\frac{g}{2}s(\be_x\bcdot\bbt^W_{1/2})\be_z \notag \\ 
&=-\frac{i\omega}{2}ge^{i\omega\tau}\varphi\ell_c'\be_z\be_x\bcdot\bW+ \frac{g}{2}se^{i\omega\tau}\varphi'\be_z\be_x\bcdot\bW+\mathrm{c.c.},
\\
\boldsymbol{\mathcal{R}}\ub{2,GG}&=\frac{g}{2}s(\be_x\bcdot\bbt^G_{1/2})\be_z=\frac{g^2}{4}s\ell_c'\be_z +\text{c.c.},\\
\boldsymbol{\mathcal{R}}\ub{3,WW}&=-\bbt^W_{1/2}\times\mathbf{F}^W_{1/2}= \bW\times\bW^*\varphi'{\varphi^*}'''+\text{c.c.},\\
\boldsymbol{\mathcal{R}}\ub{3,GW}&=-\bbt^W_{1/2}\times\mathbf{F}^G_{1/2}  -\bbt^G_{1/2}\times\mathbf{F}^W_{1/2} 
\notag \\ &=g\be_x\times\bW[\ell_c'\varphi'''-\ell_c'''\varphi']e^{i\omega\tau}+\text{c.c.},\\
\boldsymbol{\mathcal{R}}\ub{4,WW}&=\bbt^W_{1/2}\times\pd{\bbt^W_{1/2}}{s} =\bW\times\bW^*\varphi'{\varphi^*}'' + \mathrm{c.c.},\\
\boldsymbol{\mathcal{R}}\ub{4,GW}&=\bbt^W_{1/2}\times\pd{\bbt^G_{1/2}}{s} +\bbt^G_{1/2}\times\pd{\bbt^W_{1/2}}{s}\notag \\ &=ge^{i\omega\tau}\be_x\times\bW(\ell_c'\varphi''-\varphi'\ell_c'')+ \text{c.c.}
\end{align}
\end{subequations}
We note that, in simplifying some of the above expressions, we used \eqref{ellc relation}, $\bW\times\bW=\bzero$ and the fact that purely imaginary terms vanish upon adding their complex conjugates. 

The constraint \eqref{constraint transformed} gives
\begin{equation}\label{constraint 1}
\be_z\bcdot\bbt_1={\mathcal{R}}\ub{5},
\end{equation}
where we define and decompose the forcing term as
\begin{equation}
{\mathcal{R}}\ub{5}= -\frac{1}{2}\bbt_{1/2}\bcdot\bbt_{1/2} = {\mathcal{R}}\ub{5,WW}+{\mathcal{R}}\ub{5,GW}+{\mathcal{R}}\ub{5,GG}, 
\end{equation}
with
\begin{subequations}
\begin{align}
{\mathcal{R}}\ub{5,WW}&=-\frac{1}{2}\bbt^W_{1/2}\bcdot\bbt^W_{1/2}=-\frac{1}{2}\bW\bcdot\bW {\varphi'}^2e^{2i\omega\tau}-\frac{1}{2}|\bW|^2|\varphi'|^2+\mathrm{c.c.},\\
{\mathcal{R}}\ub{5,GW}&=-\bbt^W_{1/2}\bcdot\bbt^G_{1/2}=-g\bW\bcdot\be_x\varphi'\ell_c' e^{i\omega\tau}+\mathrm{c.c.},\\
{\mathcal{R}}\ub{5,GG}&=-\frac{1}{2}\bbt^G_{1/2}\bcdot\bbt^G_{1/2}=-\frac{1}{4}g^2\ell_c'^2 +  \text{c.c.}
\end{align}
\end{subequations}

The boundary conditions \eqref{bcs wall transformed} and \eqref{bcs tip transformed} give 
\begin{subequations}
\label{bcs 1}
\begin{gather}
\br_{1}=\bzero, \quad \bbt_1=\bzero \quad \text{at} \quad s=0; \tag{\theparentequation a,b} \\
\bF_1+\mathcal{F}_c\bbt_1=-\frac{1}{2}\chi\be_z+\text{c.c.}, \quad 
\quad \bM_1=\bzero \quad \text{at} \quad s=1. \tag{\theparentequation c,d}
\end{gather}
\end{subequations}

\subsubsection{Particular solution in the absence of resonance}

The homogeneous linear operator at this and all higher orders is identical to that appearing in the $\ord(\epsilon^{1/2})$ problem and in the shear-free weakly nonlinear analysis \citep{Schnitzer:25filament}. It follows that the homogeneous problem admits linear elliptical whirling solutions that are strictly horizontal and vary harmonically in $\tau$ with angular frequency $\omega$ (the fundamental harmonic). If these modes were resonantly excited by the forcing terms, the resulting secular growth in $\tau$ would invalidate the asymptotic ordering of the multiple-scale expansions \eqref{wna expansions}. In \citep{Schnitzer:25filament}, I employed an adjoint formulation to derive a ``secularity condition'' on the forcing terms that ensures that resonance is avoided. This condition at $\ord(\epsilon^{3/2})$ furnishes the amplitude equation. At the present order, however, resonance is inherently impossible: although some of the inhomogeneous terms are  fundamental harmonics (arising from bilinear interactions between the shear flow and the leading-order whirling), these forcing terms are strictly vertical, and thus geometrically orthogonal to the horizontal whirling modes. 

To proceed to the next order, we require a particular solution to the $\ord(\epsilon)$ problem. (We ignore stable modes for the same reasons as before, and note that adding complementary linear elliptical whirling solutions at this order would not affect the secularity conditions at the next order.) The form of the inhomogeneous terms suggests decomposing the particular solution as 
\begin{subequations}
\begin{align}
\br_1&=\br^{WW}_1(s,\tau,T)+\br^{GW}_1(s,\tau,T)+\br^{GG}_1(s),\\
\bbt_1&=\bbt^{WW}_1(s,\tau,T)+\bbt^{GW}_1(s,\tau,T)+\bbt^{GG}_1(s),\\
\bF_1&=-\chi\be_z+\bF^{WW}_1(s,\tau,T)+\bF^{GW}_1(s,\tau,T)+\bF^{GG}_1(s),\\
\bM_1&=\bM^{WW}_1(s,\tau,T)+\bM^{GW}_1(s,\tau,T)+\bM^{GG}_1(s).
\end{align}
\end{subequations}
Each component satisfies the inhomogeneous problem driven by the forcing terms with the matching superscript, while the explicit constant term in $\bF_1$ accounts for the constant forcing in the boundary condition (\ref{bcs 1}c). This constant term does not affect the remaining governing equations: its derivatives trivially vanish, and because it is vertical, it drops out of the $\be_z\times\bF_1$ term in the moment balance (\ref{eqs 1}c). 

The $WW$ solutions were derived in \citep{Schnitzer:25filament}:
\begin{subequations}
\label{sol1 WW}
\begin{align}
\br_1^{WW}&=-\be_z\frac{1}{2}\bW\bcdot\bW e^{2i\omega\tau}\int_0^s \{{\varphi'}(p)\}^2\,dp-\be_z\frac{1}{2}|\bW|^2\int_0^s|\varphi'(p)|^2\,dp + \text{c.c.},\\
\bbt_1^{WW}&=-\be_z\frac{1}{2}\bW\bcdot\bW {\varphi'}^2e^{2i\omega\tau}-\be_z\frac{1}{2}|\bW|^2|\varphi'|^2 + \text{c.c.},\\
\bF_1^{WW}&=\frac{1}{2}\be_z \bW\bcdot\bW e^{2i\omega\tau}\left[\mathcal{F}_c\{\varphi'(1)\}^2+\frac{i\omega}{2}\varphi^2(1)-\frac{i\omega}{2}\varphi^2(s)+i\omega\int_s^1\,\int_0^p \,\{\varphi'(q)\}^2 dq\,dp\right] \notag \\ &\quad
+\be_z\frac{\mathcal{F}_c}{2}|\bW|^2|\varphi'(1)|^2-\be_z\frac{i\omega}{2}|\bW|^2\int_s^1\varphi'(p)\varphi^*(p)\,dp +\text{c.c.},
\\
\bM_1^{WW}&=\bW\times\bW^*\varphi'{\varphi^*}'' + \text{c.c.}
\end{align}
\end{subequations}

To obtain the remaining parts of the particular solution, we anticipate, subject to \textit{a posteriori} verification, that they are vertical as are the forcing terms. In that case, constraint (\ref{constraint 1}) immediately yields
\begin{subequations}
\label{t1 GW GG}
\begin{equation}
\bbt_1^{GW}= -g\be_z(\bW\bcdot\be_x)\varphi'\ell_c' e^{i\omega\tau} + \text{c.c.}, \quad \bbt_1^{GG}=-\be_z\frac{1}{4}g^2\ell_c'^2+ \text{c.c.}
\tag{\theequation a,b}
\end{equation}
\end{subequations}
Since $\ell_c'(0)=0$ [cf.~\eqref{lc problem}], these solutions are compatible with the boundary condition (\ref{bcs 1}b) at $s=0$. Integrating (\ref{eqs 1}a) together with the boundary condition (\ref{bcs 1}a) at $s=0$, we find
\begin{subequations}
\label{r1 GW GG}
\begin{align}
\br_1^{GW}&=-ge^{i\omega\tau}\be_z(\be_x\bcdot\bW)\int_0^s\varphi'(p)\ell_c'(p)\,dp+\text{c.c.}, \\ \br_1^{GG}&=-\be_z\frac{1}{4}g^2\int_0^s\{\ell_c'(p)\}^2\,dp + \text{c.c.}
\end{align}
\end{subequations}
The constitutive relation (\ref{eqs 1}d) then gives 
\begin{subequations}
\label{M1 GW GG}
\begin{equation}
\bM_1^{GW}=ge^{i\omega\tau}\be_x\times\bW(\ell_c'\varphi''-\varphi'\ell_c'')+\text{c.c.}, \quad \bM_1^{GG}=\bzero. \tag{\theequation a,b}
\end{equation}
\end{subequations}
Since $\ell_c''(1)=0$ [cf.~\eqref{lc problem}] and $\varphi''(1)=0$ [cf.~\eqref{phi def text}], the moments \eqref{M1 GW GG} vanish at $s=1$ in accordance with the boundary condition (\ref{bcs 1}d).

To find the corresponding force contributions, we use (\ref{eqs 1}b) to obtain the derivatives 
\begin{equation}
\pd{\bF_1^{GW}}{s}=ge^{i\omega\tau}\be_z(\be_x\bcdot\bW)\left(-\frac{i\omega}{2}\int_0^s\varphi'(p)\ell_c'(p)\,dp-\frac{i\omega}{2}\varphi\ell_c'+\frac{1}{2}s\varphi'\right)+\text{c.c.}
\end{equation}
and 
\begin{equation}
\pd{\bF_1^{GG}}{s}=\frac{g^2}{4}s\ell_c'\be_z + \text{c.c.}
\end{equation}
From (\ref{bcs 1}c), these are supplemented by the boundary conditions
\begin{subequations}
\begin{align}
\bF_1^{GW}&=g\mathcal{F}_c\be_z(\be_x\bcdot\bW)\varphi'(1)\ell_c'(1)e^{i\omega\tau}+\text{c.c.} \quad \text{at} \quad s=1, \\ 
\bF_1^{GG}&=\be_z\frac{g^2\mathcal{F}_c}{4}\{\ell_c'(1)\}^2+\text{c.c.}  \quad \text{at} \quad s=1. 
\end{align}
\end{subequations}
We thus find
\begin{subequations}
\label{F1 GW GG}
\begin{align}
\bF_1^{GW}&=g\be_z(\be_x\bcdot\bW)e^{i\omega\tau}\Biggl\{\mathcal{F}_c\varphi'(1)\ell_c'(1)+\frac{i\omega}{2}\int_s^1\int_0^p\,\varphi'(q)\ell_c'(q)\,dq\,dp
\notag \\ &\quad
+\frac{i\omega}{2}\int_s^1\varphi(p)\ell_c'(p)\,dp-\frac{1}{2}\int_s^1p\varphi'(p)\,dp\Biggr\} +\text{c.c.},
\\ 
\bF_1^{GG}&=\frac{g^2}{4}\be_z\left(\mathcal{F}_c\{\ell_c'(1)\}^2-\int_s^1p\ell_c'(p)\,dp\right)+\text{c.c.}
\end{align}
\end{subequations}

Finally, it can be verified that the vector fields of the particular solution are indeed vertical, and the moment balance (\ref{eqs 1}c) is identically satisfied. 

\subsection{The $\ord(\epsilon^{3/2})$ problem}
\subsubsection{Problem formulation}
At $\ord(\epsilon^{3/2})$, the governing differential equations \eqref{pdes transformed} together with the shear-term expansion \eqref{shear expansion} give  
\begin{subequations}
\label{eqs 32}
\begin{eqnarray}
\pd{\br_{3/2}}{s}-\bbt_{3/2}&=&\bzero,  \\
 \pd{\bF_{3/2}}{s}-\left(\tI-\frac{1}{2}\be_z\be_z\right)\bcdot \pd{\br_{3/2}}{\tau}&=&\boldsymbol{\mathcal{S}}\ub{2}, \\ 
\pd{\bM_{3/2}}{s}+\mathcal{F}_c\be_z\times\bbt_{3/2}+\be_z\times\bF_{3/2}&=&\boldsymbol{\mathcal{S}}\ub{3}, \\ 
\bM_{3/2}-\be_z\times\pd{\bbt_{3/2}}{s}&=&\boldsymbol{\mathcal{S}}\ub{4}.
\end{eqnarray}
\end{subequations}
Noting that the $\ord(\epsilon^{1/2})$ fields are horizontal, while the $\ord(\epsilon)$ ones are vertical, the new forcing terms are found as
\begin{subequations}
\begin{align}
\boldsymbol{\mathcal{S}}\ub{2} &= \pd{\br_{1/2}}{T}-\frac{1}{2}\bbt_{1/2}\be_z\bcdot\pd{\br_{1}}{\tau}  - \frac{1}{2}\bbt_{1/2}\bbt_{1/2}\bcdot\pd{\br_{1/2}}{\tau} \notag \\ &\quad+g\left[-z_1\be_x+\frac{1}{2}s(\be_x\bcdot\bbt_{1/2})\bbt_{1/2}\right],
\\
\boldsymbol{\mathcal{S}}\ub{3} &= -\bbt_{1/2}\times\bF_{1}-\bbt_1\times\bF_{1/2},\\
\boldsymbol{\mathcal{S}}\ub{4}&=\bbt_{1/2}\times\pd{\bbt_1}{s}+\bbt_1\times\pd{\bbt_{1/2}}{s}.
\end{align}
\end{subequations}
 
Furthermore, the constraint \eqref{constraint transformed} gives (noting that $\bt_1\bcdot\bt_{1/2}=0$) 
\begin{equation}\label{constraint 3/2}
\bt_{3/2}\bcdot\be_z=0,
\end{equation}
whereas the boundary conditions \eqref{bcs wall transformed} and \eqref{bcs tip transformed} give 
\begin{subequations}
\label{bcs 32}
\begin{gather}
\br_{3/2}=\bzero, \quad \bbt_{3/2}=\bzero \quad \text{at} \quad s=0; \tag{\theparentequation a,b}\\
\bF_{3/2}+\mathcal{F}_c\bbt_{3/2}=-\chi\bbt_{1/2}^W-\chi\bbt_{1/2}^G, \quad \bM_{3/2}=\bzero \quad \text{at} \quad s=1. \tag{\theparentequation c,d}
\end{gather}
\end{subequations}
 
\subsubsection{Elimination of secular terms} 
At the present order, the forcing terms are purely horizontal and include fundamental harmonics. Depending also on their spatial dependence, these terms may therefore resonantly excite linear elliptical whirling modes leading to a breakdown of regularity of the weakly nonlinear expansions \eqref{wna expansions}. Eliminating those resonant forcing terms furnishes the requisite amplitude equation governing $\bW(T)$. 

As a first step, we symbolically isolate the fundamental harmonics in the forcing terms, collecting all other terms under the label ``n.s.t.,'' which stands for nonsecular terms. To this end, we write the forcing terms appearing in the differential equations \eqref{eqs 32} as
\begin{equation}\label{fk def}
\boldsymbol{\mathcal{S}}\ub{k}=\{e^{i\omega\tau}\bbf\ub{k}(s) + \text{c.c.}\} + \text{n.s.t.}, \quad k=2,3,4.
\end{equation}
Similarly, we decompose the right-hand sides of the boundary condition (\ref{bcs 32}c) as
\begin{equation}
e^{i\omega\tau}\mathbf{h} + \text{c.c.} + \text{n.s.t.},
\end{equation}
where, since the shear term in that boundary condition is constant in $\tau$, 
\begin{equation}\label{h forcing}
\mathbf{h}=-\chi \bW\varphi'(1). 
\end{equation}

We may now directly invoke the secularity condition derived in \citep{Schnitzer:25filament}: 
\begin{equation}\label{solvability}
\int_0^1\left({\phi^*}'\bbf\ub{2}-{\phi^*}''\be_z\times\bbf\ub{3}-{\phi^*}'''\be_z\times\bbf\ub{4}\right)\,ds={\phi^*}'(1)\mathbf{h},
\end{equation}
wherein $\phi(s)$ is an adjoint eigenfunction satisfying 
\begin{equation}\label{adjoint def}
\phi''''+\mathcal{F}_c\phi''-i\omega \phi=0, \quad \phi'(0)=\phi''(0)=\phi(1)=\phi'''(1)+\mathcal{F}_c\phi'(1)=0. 
\end{equation}
This secularity condition is independent of the normalization of $\phi(s)$; we find it convenient to fix it by imposing $\phi(0)=1$.

To extract $\bbf\ub{k}(s)$, $k=2,3,4$, we note the decompositions
\begin{subequations}
\label{S terms}
\begin{align}
\boldsymbol{\mathcal{S}}\ub{2}&=\left(\tI-\frac{1}{2}\be_z\be_z\right)\bcdot \pd{\br^W_{1/2}}{T}+\boldsymbol{\mathcal{S}}\ub{2,WWW}+\boldsymbol{\mathcal{S}}\ub{2,GWW}
\notag \\ &\quad +\boldsymbol{\mathcal{S}}\ub{2,GGW}+\boldsymbol{\mathcal{S}}\ub{2,GGG},
\\
\boldsymbol{\mathcal{S}}\ub{3} 
&= -\chi\be_z\times\bbt_{1/2}^W-\chi\be_z\times\bbt_{1/2}^G
+\boldsymbol{\mathcal{S}}\ub{3,WWW}+\boldsymbol{\mathcal{S}}\ub{3,GWW}
\notag \\ &\quad +\boldsymbol{\mathcal{S}}\ub{3,GGW}+\boldsymbol{\mathcal{S}}\ub{3,GGG},
\\
\boldsymbol{\mathcal{S}}\ub{4}&=\boldsymbol{\mathcal{S}}\ub{4,WWW}+\boldsymbol{\mathcal{S}}\ub{4,GWW}+\boldsymbol{\mathcal{S}}\ub{4,GGW}+\boldsymbol{\mathcal{S}}\ub{4,GGG}.
\end{align}
\end{subequations}
The first terms in (\ref{S terms}a) and (\ref{S terms}b) are fundamental harmonics. The second term in (\ref{S terms}b) is steady and can therefore be discarded. 
The $WWW$ terms represent cubic interactions of the leading-order whirling; these include fundamental harmonics (as well as third harmonics). The $GGW$ terms represent quadratic interactions of the steady shear, further interacting with the whirling; these terms are therefore fundamental harmonics. The $GWW$ terms represent quadratic interactions of the whirling, further interacting with the steady shear; these terms accordingly contain zeroth and second harmonics and can therefore be discarded. Last, the $GGG$ terms represent cubic interactions of the steady shear and can therefore be discarded.

The relevant cubic terms can be expressed as 
\begin{subequations}
\begin{align}
\boldsymbol{\mathcal{S}}\ub{2,WWW}&=-\frac{1}{2}\bbt_{1/2}^W\pd{z_1^{WW}}{\tau}-\frac{1}{2}\bbt_{1/2}^W\bbt_{1/2}^W\bcdot\pd{\br_{1/2}^W}{\tau},\\
\boldsymbol{\mathcal{S}}\ub{2,GGW}&=-\frac{1}{2}\bbt_{1/2}^G\pd{z_1^{GW}}{\tau}-\frac{1}{2}\bbt_{1/2}^G\bbt_{1/2}^G\bcdot\pd{\br_{1/2}^W}{\tau}-gz_1^{GW}\be_x \notag \\ &\quad +\frac{1}{2}gs(\be_x\bcdot\bbt_{1/2}^W)\bbt_{1/2}^G+\frac{1}{2}gs(\be_x\bcdot\bbt_{1/2}^G)\bbt_{1/2}^W,\\
\boldsymbol{\mathcal{S}}\ub{3,WWW}&=-\bbt_{1/2}^W\times\bF_1^{WW}-\bbt_1^{WW}\times\bF_{1/2}^W,\\\boldsymbol{\mathcal{S}}\ub{3,GGW}&=-\bbt_{1/2}^W\times\bF_1^{GG}-\bbt_{1/2}^G\times\bF_1^{GW}-\bbt_1^{GW}\times\bF_{1/2}^G-\bbt_1^{GG}\times\bF_{1/2}^W,\\
\boldsymbol{\mathcal{S}}\ub{4,WWW}&=\bbt_{1/2}^W\times\pd{\bbt_1^{WW}}{s}+\bbt_1^{WW}\times\pd{\bbt_{1/2}^W}{s},\\
\boldsymbol{\mathcal{S}}\ub{4,GGW}&=\bbt_{1/2}^W\times\pd{\bbt_1^{GG}}{s}+\bbt_{1/2}^G\times\pd{\bbt_1^{GW}}{s}+\bbt_1^{GW}\times\pd{\bbt_{1/2}^G}{s}+\bbt_1^{GG}\times\pd{\bbt_{1/2}^W}{s}.
\end{align}
\end{subequations}
 
Substituting the $\ord(\epsilon^{1/2})$ solutions \eqref{sol 1/2} and \eqref{sol 1/2 G}, and the $\ord(\epsilon)$ solutions \eqref{sol1 WW}, \eqref{r1 GW GG}, \eqref{t1 GW GG}, \eqref{M1 GW GG} and \eqref{F1 GW GG}, allows us to identify $\bbf\ub{k}(s)$ in a form that isolates the shear-induced contributions,
\begin{equation}
\label{f decomp}
\bbf\ub{k}(s)=\bar{\bbf}\ub{k}(s)+g^2\Delta\bbf\ub{k}(s), \quad k=2,3,4,
\end{equation}
where $\bar{\bbf}\ub{k}(s)$ and $\Delta\bbf\ub{k}(s)$ are $g$-independent, with $\bar{\bbf}\ub{k}(s)$ identical to the corresponding forcing terms present in the shear-free analysis \citep{Schnitzer:25filament}: 
\begin{subequations}
\label{f W}
\begin{align}
\bar{\bbf}\ub{2}&=\frac{d\bW}{dT}\varphi+\frac{i\omega}{2}\bW^*(\bW\bcdot\bW){\varphi^*}'\int_0^s\{\varphi'(p)\}^2\,dp \notag \\
&\quad -\frac{i\omega}{2}\left[\bW|\bW|^2\varphi'({\varphi^*}'\varphi-\varphi'{\varphi^*})+\bW^*(\bW\bcdot\bW)\varphi|\varphi'|^2\right],
\\
\bar{\bbf}\ub{3}&=
-\be_z\times\bW\varphi'\chi \notag \\
&\quad +\be_z\times\bW|\bW|^2\left[\mathcal{F}_c\varphi'(|\varphi'(1)|^2-|\varphi'|^2)-|\varphi'|^2\varphi'''+\omega\varphi'\mathrm{Im}\int_s^1\varphi'(p)\varphi^*(p)\,dp\right] \notag \\
&\quad +\frac{1}{2}\be_z\times\bW^*(\bW\bcdot\bW)\Biggl[-{\varphi'}^2(\varphi'''+\mathcal{F}_c\varphi')^*+\mathcal{F}_c{\varphi^*}'\{\varphi'(1)\}^2+\frac{i\omega}{2}{\varphi^*}'\varphi^2(1) \notag \\ &\quad -\frac{i\omega}{2}{\varphi^*}'\varphi^2+i\omega{\varphi^*}'\int_s^1\int_0^p \{\varphi'(q)\}^2dq\,dp\Biggr],
\\
 \bar{\bbf}\ub{4}&=\be_z\times\bW |\bW|^2{\varphi'}^2{\varphi^*}''   + \be_z\times\bW^*(\bW\bcdot\bW)\left(\varphi''|\varphi'|^2-\frac{1}{2}{\varphi'}^2{\varphi^*}''\right).
\end{align}
\end{subequations}
The new shear-induced terms $\Delta\bbf\ub{k}(s)$ are found as
\begin{subequations}\label{f new}
\begin{align}
\Delta\bbf\ub{2}&=\be_x\be_x\bcdot\bW\left\{\left(1+\frac{i\omega}{2}\ell_c'\right)\int_0^s\varphi'(p)\ell_c'(p)\,dp+\frac{1}{2}s\varphi'\ell_c'-\frac{i\omega}{2}\ell_c'^2\varphi\right\} \notag \\ &\quad +\frac{1}{2}\bW s\ell_c'\varphi',\\
\Delta\bbf\ub{3} &=\frac{1}{2}\be_z\times\bW\left[\varphi'\left(\mathcal{F}_c\{\ell_c'(1)\}^2-\int_s^1p\ell_c'(p)\,dp\right)-\ell_c'^2(\varphi'''+\mathcal{F}_c\varphi')\right] \notag \\
&\quad +\be_y(\be_x\bcdot\bW)\Biggl\{ \ell_c' \Biggl[\mathcal{F}_c\varphi'(1)\ell_c'(1)+\frac{i\omega}{2}\int_s^1\int_0^p\,\varphi'(q)\ell_c'(q)\,dq\,dp
\notag \\
&\quad +\frac{i\omega}{2}\int_s^1\varphi(p)\ell_c'(p)\,dp-\frac{1}{2}\int_s^1p\varphi'(p)\,dp\Biggr] +\frac{1}{2}\varphi'\ell_c'[1-s^2-2\mathcal{F}_c\ell_c'(1)]\Biggr\},
\\
\Delta\bbf\ub{4}&=
(\be_z\times\bW)(\ell_c'\ell_c''\varphi'-\frac{1}{2}\ell_c'^2\varphi'')+\be_y(\be_x\bcdot\bW)\ell_c'^2\varphi''.
\end{align}
\end{subequations}

\subsubsection{Amplitude equation}
It remains only to substitute the forcing terms \eqref{h forcing} and \eqref{f decomp} into the secularity condition \eqref{solvability} and evaluate the resulting quadratures. Upon using \eqref{f W} and \eqref{f new} for $\bar{\bbf}\ub{k}$ and $\Delta\bbf\ub{k}$, respectively, this yields the  amplitude equation
\begin{equation}\label{amplitude W}
\frac{d\bW}{dT}=\alpha\bW^*\bW\bcdot\bW+\beta\bW\bW^*\bcdot\bW+[\gamma\chi-g^2\xi^{\parallel}]\be_x\be_x\bcdot\bW+(\gamma\chi-g^2\xi^{\perp})\be_y\be_y\bcdot\bW,
\end{equation}
where $\alpha$, $\beta$, $\gamma$, $\xi^{\parallel}$ and $\xi^{\perp}$ are complex constants defined through quadratures involving the eigenfunction $\varphi(s)$ [cf.~\eqref{phi def text}], the adjoint eigenfunction $\phi(s)$ [cf.~\eqref{adjoint def}], and, in the case of $\xi^{\parallel}$ and $\xi^{\perp}$, also $\ell_c(s)$ [cf.~\eqref{ell solution app}].  

The constants $\alpha$, $\beta$, $\gamma$ appear also in the shear-free amplitude equation derived in \citep{Schnitzer:25filament}. Upon defining the integral 
\begin{equation}
\Lambda = -\int_0^1{\phi^*}'\varphi\,ds,
\end{equation}
these constants are given as
\begin{subequations}
\label{alpha beta gamma quad}
\begin{align}
\Lambda\alpha &= \frac{i\omega}{2}\left\{\int_0^1{\phi^*}'(s){\varphi^*}'(s)\int_0^s \{\varphi'(p)\}^2dp \,ds-\int_0^1{\phi^*}'\varphi|\varphi'|^2\,ds\right\} \notag \\
&\quad +\frac{1}{2}\int_0^1{\phi^*}''\Biggl\{-{\varphi'}^2(\varphi'''+\mathcal{F}_c\varphi')^*+\{\varphi'(1)\}^2\mathcal{F}_c{\varphi^*}'+\frac{i\omega}{2}\varphi^2(1){\varphi^*}' \notag \\
&\quad -\frac{i\omega}{2}{\varphi^*}'\varphi^2+i\omega{\varphi^*}'\int_s^1 \int_0^p  \{\varphi'(q)\}^2dq\,dp\Biggr\}ds\notag \\ &\quad +\int_0^1{\phi^*}'''(\varphi''|\varphi'|^2-\frac{1}{2}{\varphi'}^2{\varphi^*}'')\,ds,
\\
\Lambda \beta &= \int_0^1{\phi^*}'''\varphi'^2{\varphi^*}''\,ds -\frac{i\omega}{2}\int_0^1{\phi^*}'\varphi'({\varphi^*}'\varphi-\varphi'{\varphi^*})\,ds \notag \\&\quad + 
\int_0^1{\phi^*}''\Biggl\{-|\varphi'|^2(\varphi'''+\mathcal{F}_c\varphi') +\mathcal{F}_c|\varphi'(1)|^2\varphi' \notag \\ 
&\quad +
 \frac{i\omega}{2}\varphi'\left[|\varphi(1)|^2-|\varphi|^2-2\int_s^1\varphi'(p){\varphi^*}(p)\,dp\right]\Biggr\}\,ds,
\\
\Lambda\gamma &= \int_0^1{\phi^*}'\varphi''\,ds.
\end{align}
\end{subequations}

The new constants $\xi^\parallel$ and $\xi^{\perp}$ are found as 
\begin{subequations}
\label{xi quad}
\begin{align}
-\Lambda \xi^\parallel&= \int_0^1{\phi^*}'\left[\left(1+\frac{i\omega}{2}\ell_c'\right)\int_0^s\varphi'(p)\ell_c'(p)\,dp+s\varphi'\ell_c'-\frac{i\omega}{2}\ell_c'^2\varphi\right]\,ds
\notag \\
&\quad +
\frac{1}{2}\int_0^1 {\phi^*}''\left[\varphi'\left(\mathcal{F}_c\{\ell_c'(1)\}^2-\int_s^1p\ell_c'(p)\,dp\right)-\ell_c'^2(\varphi'''+\mathcal{F}_c\varphi')\right]\,ds \notag \\
&\quad +\int_0^1  {\phi^*}''\Biggl\{ \ell_c' \Biggl[\mathcal{F}_c\varphi'(1)\ell_c'(1)+\frac{i\omega}{2}\int_s^1\int_0^p\varphi'(q)\ell_c'(q)\,dq\,dp \notag \\ &\quad
+\frac{i\omega}{2}\int_s^1\varphi(p)\ell_c'(p)\,dp-\frac{1}{2}\int_s^1p\varphi'(p)\,dp\Biggr] +\frac{1}{2}\varphi'\ell_c'[1-s^2-2\mathcal{F}_c\ell_c'(1)]\Biggr\}\,ds \notag \\ 
&\quad + \int_0^1{\phi^*}'''\left(\ell_c'\ell_c''\varphi'+\frac{1}{2}\ell_c'^2\varphi''\right)\,ds,
\\ 
-\Lambda\xi^{\perp}&= \frac{1}{2}\int_0^1{\phi^*}'s\ell_c'\varphi'\,ds \notag \\
&\quad + 
\frac{1}{2}\int_0^1 {\phi^*}''\left[\varphi'\left(\mathcal{F}_c\{\ell_c'(1)\}^2-\int_s^1p\ell_c'(p)\,dp\right)-\ell_c'^2(\varphi'''+\mathcal{F}_c\varphi')\right]\,ds
\notag \\
&\quad + \int_0^1{\phi^*}'''\left(\ell_c'\ell_c''\varphi'-\frac{1}{2}\ell_c'^2\varphi''\right)\,ds.
\end{align}
\end{subequations}

Recalling that $\bA\sim \epsilon^{1/2}\bW$, $\mathcal{G}=\epsilon^{1/2}g$ and $\Delta\mathcal{F}=\epsilon \chi$, and that $t=T/\epsilon$ on the physical diagonal \eqref{diagonal}, the amplitude equation \eqref{amplitude W} becomes \eqref{amplitude shear} in the main text. The quadratures \eqref{alpha beta gamma quad} and \eqref{xi quad} are evaluated in Mathematica (Supplementary Information, Part I) using semi-analytical solutions for the eigenfunctions $\varphi(s)$ and $\phi(s)$---incorporating numerical solutions for the threshold values $\mathcal{F}_c$ and $\omega$---and the analytical solution for $\ell_c(s)$. This yields the values \eqref{coefficients} given in the main text. Furthermore, the AI-assisted  Mathematica code mentioned at the beginning of this Appendix (Supplementary Information, Part II) provides an independent reproduction of these coefficients.

\end{appendices}

\bibliography{refs.bib}

\end{document}